\documentclass[notitlepage]{revtex4-1}

\usepackage[utf8]{inputenc}
\usepackage[colorlinks=true,citecolor=blue,linkcolor=blue]{hyperref}
\usepackage[normalem]{ulem}
\usepackage{url}
\usepackage{graphicx,wrapfig,float,slashed,cancel}
\usepackage{amsmath,amssymb,epsfig,graphicx,xcolor,stmaryrd}
\usepackage{bm}
\usepackage{enumitem}
\usepackage{orcidlink}

\begin{document}

\title{Properties of doubly heavy spin-$\frac{1}{2}$ baryons: The ground and excited states}

	\author{M.~Shekari Tousi$^{a}$\orcidlink{0009-0007-7195-0838}}
    \email{marzie.sh.tousi@ut.ac.ir }
	
	\author{K.~Azizi$^{a,b}$\orcidlink{0000-0003-3741-2167}} 
	 \email{kazem.azizi@ut.ac.ir} \thanks{Corresponding author} 
	
	\affiliation{
		$^{a}$Department of Physics, University of Tehran, North Karegar Avenue, Tehran 14395-547, Iran\\
		$^{b}$Department of Physics, Do\v{g}u\c{s} University, Dudullu-\"{U}mraniye, 34775
		Istanbul, Turkey
	}
	
\date{\today}

\preprint{}

\begin{abstract}

We determine the masses and residues of the ground and excited spin-$\frac {1}{2} $ baryons consist of two heavy b or c quark utilizing the QCD sum rule formalism. In the calculations, we consider the nonperturbative operators up to ten mass dimensions in order to increase the accuracy compared to the previous calculations. We report the obtained results for both the symmetric and antisymmetric currents defining the doubly heavy baryons of the ground state (1S), first orbitally excited state (1P) and first radially excited state (2S). We compare our results with the predictions of other nonperturbative approaches as well as existing experimental data which is available only for the ground state of $	\Xi_{cc}$ channel. These predictions can help the experimental groups in their searches for all members of the doubly heavy baryons in their ground and exited states.

\end{abstract}


\maketitle

\renewcommand{\thefootnote}{\#\arabic{footnote}}
\setcounter{footnote}{0}
\section{\label{sec:level1}Introduction}\label{intro}

Investigation of the properties of the doubly heavy baryons is an important and useful area in particle physics. The quark model makes predictions about the hadronic states containing single, double and triple heavy quarks~\cite{GellMann:1964nj}. All of the hadronic states with a single heavy quark predicted by the quark model have already been observed in experiments. 
 For a while, scientists were unable to observe baryons consisting of two heavy quarks, posing a longstanding puzzle within the quark model. However, a breakthrough was made in 2002 by the SELEX collaboration with the discovery of $\Xi^+_{cc}(3520)$ in the $ p D^+ K^-$ decay channel \cite{SELEX:2002wqn}. This discovery was later confirmed by the same experiment in 2005 \cite{SELEX:2004lln}. In 2012, a calculation showed that the mass of $\Xi_{cc}$ state may be greater than the one measured by SELEX collaboration \cite{Aliev:2012ru}. In 2017, the LHCb collaboration announced the discovery of $\Xi^{++}_{cc}(3621)$ through the $ \Xi^{++}_{cc}\rightarrow \Lambda^+_{c} K^- \pi^+ \pi^+$decay channel \cite{LHCb:2017iph}, which was later confirmed by the same collaboration in 2018 with a statistical significance of 5.9 $\sigma$ in the decay channel $ \Xi^{++}_{cc}\rightarrow \Xi^+_{c} \pi^+$ \cite{LHCb:2018pcs}. As it is seen, the mass of $\Xi^{++}_{cc}$ measured by LHCb was about 100 $\mathrm {MeV}$ greater than the measured mass for the $\Xi^+_{cc}$ by SELEX, representing a puzzle since a unit charge difference cannot produce such a large mass difference. Hence this subject became one of the hot topics in hadron physics both from the experimental and theoretical views. Following this,  in 2019, a search for $\Xi^+_{cc}$ was  performed through $ \Xi^{+}_{cc}\rightarrow \Lambda^+_{c} K^- \pi^+$ decay channel  also by the LHCb collaboration \cite{LHCb:2019gqy}. The results of this search were combined with the results of another search that was conducted in 2021 on the decay channel $ \Xi^{+}_{cc}\rightarrow \Xi^{+}_{c} \pi^- \pi^+ $, leading to a maximum local significance of 4.0 standard deviations around
 	the mass of 3620 $ \mathrm {MeV}$ for the $\Xi^{+}_{c}$, including systematic uncertainties \cite{LHCb:2021eaf}.  This value does not exhibit such a difference compared to $ \Xi_{cc}^{++} $’s mass. Experimental groups also continued to discover other members  of these types of baryons, but so far it has not led to the discovery of another new doubly heavy baryons (for instance see Ref. \cite{LHCb:2021xba}). Numerous theoretical researches have been conducted based on these discoveries to determine the properties of the doubly heavy baryons such as mass, lifetime, residue, decay widths, etc. within various methods \cite{Ebert:2002ig,Zhang:2008rt,Wang:2010hs,Rostami:2020euc,Aliev:2021hqq,Azizi:2020zin,Olamaei:2020bvw,Yu:2017zst,Luchinsky:2020fdf,Gerasimov:2019jwp,Wang:2017mqp,Meng:2017udf,Gutsche:2017hux,Xiao:2017udy,Lu:2017meb,Xiao:2017dly,Zhao:2018mrg,Xing:2018lre,Jiang:2018oak,Gutsche:2018msz,Gutsche:2019wgu,Yu:2019yfr,Gutsche:2019iac,Ke:2019lcf,Cheng:2020wmk,Hu:2020mxk,Li:2020qrh,Han:2021gkl,Wang:2017azm,Shi:2017dto,Zhang:2018llc,Ivanov:2020xmw,Rahmani:2020pol,Li:2017pxa,Berezhnoy:2018bde,Guo:2017vcf,Ma:2017nik,Yao:2018zze,Yao:2018ifh,Meng:2018zbl,Shi:2020qde,Qiu:2020omj,Olamaei:2021hjd,Qin:2021dqo,Hu:2017dzi,Li:2018epz,Shi:2019hbf,Aliyev:2022rrf,Shi:2019fph,Alrebdi:2020rev,Aliev:2020aon,Aliev:2012iv,Aliev:2019lvd,Aliev:2012ru,Aliev:2012nn,Ozdem:2018uue,Ozdem:2019zis,Padmanath:2019ybu,Brown:2014ena,Giannuzzi:2009gh,Shah:2017liu,Shah:2016vmd,Yoshida:2015tia,Valcarce:2008dr,Wang:2010it,Ortiz-Pacheco:2023kjn,Sharma:2017txj,Wang:2018lhz}. Nevertheless, triply heavy baryons are still waiting to be found in the future experiments.
 
 In this study, we calculate the masses and residues of the doubly heavy spin-$\frac {1}{2} $ baryons in their ground and exited states. For calculation of these parameters, some nonperturbative approaches are needed. One of these methods is the QCD sum rule formalism which was introduced in 1979 by Shifman, Vainshtein and Zakharov based on the fundamental QCD Lagrangian, considering the correlation function including different interpolating currents \cite{Shifman:1978bx, Shifman:1978by}. This method is a successful approach in hadron physics and has many applications in the study of hadrons \cite{Aliev:2010uy,Aliev:2009jt,Aliev:2012ru,Agaev:2016dev,Azizi:2016dhy} confirmed by different experiments. In various approaches, like lattice QCD \cite{Padmanath:2019ybu,Brown:2014ena}, relativistic quark model \cite{Ebert:2002ig}, quark model \cite{Yoshida:2015tia}, hypercentral constituent quark model \cite{Shah:2016vmd,Shah:2017liu}, Faddeev method \cite{Valcarce:2008dr}, Regge phenomenology \cite{Oudichhya:2023pkg}, Hamiltonian model \cite{Yoshida:2015tia}, QCD sum rules \cite{Aliev:2019lvd,Zhang:2008rt,Wang:2010it,Wang:2010hs,Aliev:2012ru,Aliev:2012iv}, Bathe-Salpeter equations \cite{Li:2022ywz}, Salpeter model \cite{Giannuzzi:2009gh} and hypercentral constituent quark model \cite{Shah:2016vmd,Shah:2017liu}, researchers studied the spectroscopic parameters of the doubly heavy baryons. To investigate any type of interaction/decay of the doubly heavy baryons, we need their exact value of the mass and residue. For this purpose, we compute the spectroscopic parameters with the aim of increasing the accuracy in the calculations by applying the higher dimensional nonperturbative effects in the ground, first orbitally and radially excited states.

Calculations of the mass of the doubly heavy baryons in the ground state have been done in different approaches such as relativistic quark model \cite{Ebert:2002ig}, double ratios of sum rules (DRSR) \cite{Narison:2010py} and QCD sum rule method \cite{Lichtenberg:1995kg,Zhang:2008rt,Aliev:2012ru,Wang:2010hs}. With the aim of reducing the error and increasing the accuracy by including the higher-dimensional nonperturbative operators, we perform these calculations up to 10 dimensions. In the previous work done by sum rules \cite{Aliev:2012ru}, the calculations were done up to 5 mass dimensions and the reported error values are relatively high.

 Calculations of the mass of the doubly heavy baryons for $1P$ excited state have also been done in various methods such as relativistic quark model \cite{Ebert:2002ig}, quark model \cite{Roberts:2007ni,Yoshida:2015tia}, Faddeev method \cite{Valcarce:2008dr} and QCD sum rules \cite{Aliev:2019lvd,Wang:2010it}. These calculations for $2S$ excited state have also been done in different approaches such as relativistic quark model \cite{Ebert:2002ig}, quark model \cite{Yoshida:2015tia}, Salpeter model \cite{Giannuzzi:2009gh}, hypercentral constituent quark model \cite{Shah:2016vmd,Shah:2017liu}, Faddeev method \cite{Valcarce:2008dr} and QCD sum rule formalism \cite{Aliev:2019lvd}. For the $1P$ and $2S$ excited states, we calculate the mass of all the members more accurately by including the higher dimensional nonperturbative operators as discussed above.
 
 The calculation of the residues of these baryons has been performed for the ground states and their first radial and orbital excitations in fewer references. For the ground states, researchers studied this parameter using QCD sum rule method \cite{Zhang:2008rt,Aliev:2012ru,Wang:2010hs}. These studies for the $1P$ excited state have been done in QCD sum rules \cite{Aliev:2019lvd,Wang:2010hs}. For the $2S$ excited state the residue calculation has also been done in Ref. \cite{Aliev:2019lvd}. We report the results of our calculations for the residue of these baryons in the ground, $1P$ and $2S$ states with the higher accuracy in the operator product expansion (OPE) as well. The obtained results can be used in the future experiments of particle laboratories such as LHC, to research for these types of baryons.

This work is organized as follows: Section II presents the derivation of sum rules for masses and residues of the doubly heavy baryons, while Sec. III reports the results of numerical analysis of the sum rules and comparison of our findings with the existing literature. Section IV includes summary and conclusion and we move some expressions to the appendix.

\section{Calculation of the mass and residue of the doubly heavy baryons}\label{II}

The QCD sum rule method follows a standard prescription, where a correlation function is evaluated through two distinct approaches. The first approach involves the use of hadronic degrees of freedom, referred to as the physical or hadronic side. This approach yields results that contain physical quantities like the mass and residue of the considered states. The second approach involves QCD degrees of freedom, such as quark-gluon condensates, QCD coupling constant and quark masses. This approach gives the QCD side of the calculations. QCD sum rules for the physical quantities are obtained by matching the results of both sides and considering the coefficients of the same Lorentz structures. To start the calculation, we need to consider the two-point correlation function as following form:
\begin{equation}
	\Pi(q)=i\int d^{4}xe^{iq\cdot x}\langle 0|\mathcal{T}\{\eta(x)\bar{\eta}(0)\}|0\rangle ,\label{eq:CorrF1}
\end{equation}        
where $\eta(x)$ represents the interpolating current of the doubly heavy baryons and $\mathcal{T}$ is the time ordering operator. The interpolating current of symmetric and antisymmetric (concerning the heavy quark exchange) of these baryons are
\begin{align}\label{eq:CorrF2}
	\eta^S=\frac{1}{\sqrt{2}} \epsilon_{abc} \{(Q^{a^T} C q^b)\gamma_5 Q'^c  +(Q'^{a^T} C q^b)\gamma_5 Q^c +t (Q^{a^T} C \gamma_5 q^b) Q'^c+t(Q'^{a^T} C \gamma_5 q^b) Q^c\},
\end{align}

and
\begin{align}\label{eq:CorrF3}
	\eta^A&=\frac{1}{\sqrt{6}} \epsilon_{abc} \{2(Q^{a^T} C Q'^b)\gamma_5 q^c +(Q^{a^T} C q^b)\gamma_5 Q'^c  -(Q'^{a^T} C q^b)\gamma_5 Q^c  +2t (Q^{a^T} C \gamma_5 Q'^b) q^c+t (Q^{a^T} C \gamma_5 q^b) Q'^c  \nonumber\\
	&-t (Q'^{a^T} C \gamma_5 q^b) Q^c\},
\end{align}
where $Q$ and $Q'$ represents heavy quark field ($b$ or $c$); $a$, $b$ and $c$ are color indices, $C$ is the charge conjugation operator and $t$ is an arbitrary mixing parameter to be fixed from the analysis. $t=-1$ corresponds to the Ioffe current. The above interpolating currents are written considering all the quantum numbers of the states under study. It should be noted that the doubly heavy baryons including $	\Xi_{cc}$, $	\Xi_{bc}$, $	\Xi_{bb}$, $	\Omega_{cc}$, $	\Omega_{bc}$ and $	\Omega_{bb}$ are symmetric and $	\Xi^{\prime}_{bc}$ and $	\Omega^{\prime}_{bc}$ are antisymmetric with respect to the exchange of the two heavy quarks.

For the physical side, the correlator is transformed into its final form by inserting complete sets of hadronic states in the relevant places:

\begin{align}
	\Pi^{\mathrm{Phys}}(q)&=\frac{\langle0|\eta|B_{QQ'}(q,s)\rangle\langle B_{QQ'}(q,s)|\bar{\eta}|0\rangle}{m^2-q^2}
	+\frac{\langle0|\eta|\tilde{B}_{QQ'}(q,s)\rangle\langle\tilde{B}_{QQ'}(q,s)|\bar{\eta}|0\rangle}{\tilde{m}^2-q^2}
	+\frac{\langle0|\eta| B_{QQ'}'(q,s)\rangle\langle B_{QQ'}'(q,s)|\bar{\eta}|0\rangle}{m'{}^2-q^2}\nonumber\\
	& +\cdots.
	\label{Eq:cor:Phys}
\end{align}
The $|B_{QQ'}(q,s)\rangle$, $|\tilde{B}_{QQ'}(q,s)\rangle$ and $|B_{QQ'}'(q,s)\rangle$ are the one-particle states of the ground (positive parity), first orbital excitation $1P$ (negative parity) and first radial excitation $2S$ (positive parity) states of baryon, respectively. Here, $m$, $\tilde{m}$ and $m'$ are their corresponding masses and $\cdots$ represents the contributions of the higher states and continuum. The matrix elements in Eq.~(\ref{Eq:cor:Phys}) are defined as follows:
\begin{eqnarray}
	\langle 0|\eta|B_{QQ'}(q,s)\rangle&=&\lambda u(q,s),\nonumber\\
	\langle 0|\eta|\tilde{B}_{QQ'}(q,s)\rangle&=&\tilde{\lambda}\gamma_5 u(q,s),\nonumber\\
	\langle 0|\eta|B_{QQ'}'(q,s)\rangle&=&\lambda' u(q,s),
\end{eqnarray}  
where $\lambda$, $\tilde{\lambda}$ and $\lambda'$ are the corresponding residues and $u(q,s)$ is the Dirac spinor. These matrix elements are used in Eq.~(\ref{Eq:cor:Phys}) and summation over spins of Dirac spinors, which is given as
\begin{eqnarray}
	\sum_{s}u(q,s)\bar{u}(q,s)=(\not\!q+m),
\end{eqnarray}  
is performed. So, the physical side becomes
\begin{eqnarray}
	\Pi^{\mathrm{Phys}}(q)=\frac{\lambda^2(\not\!q+m)}{m^2-q^2}+\frac{\tilde{\lambda}^2(\not\!q-\tilde{m})}{\tilde{m}^2-q^2}+\frac{\lambda'^2(\not\!q+m')}{m'{}^2-q^2}+\cdots.
	\label{Eq:cor:Phys1}
\end{eqnarray}
After the Borel transformation, the final result for the physical side becomes
\begin{eqnarray}
	\tilde{\Pi}^{\mathrm{Phys}}(q)=\lambda^2(\not\!q+m)e^{-\frac{m^2}{M^2}}+\tilde{\lambda}^2(\not\!q-\tilde{m})e^{-\frac{\tilde{m}^2}{M^2}}+\lambda'^2(\not\!q+m')e^{-\frac{m'{}^2}{M^2}}+\cdots,
	\label{Eq:cor:Fin}
\end{eqnarray}
where $\tilde{\Pi}^{\mathrm{Phys}}(q)$ denotes the correlation function after the Borel transformation.

To analyze the QCD side, we use Eq.~(\ref{eq:CorrF1}) and substitute the interpolating currents from Eqs.~(\ref{eq:CorrF2}) and (\ref{eq:CorrF3}) into it. The calculations involve conducting possible contractions among the quark fields utilizing Wick's theorem. The contracted quark fields are substituted with heavy and light quarks propagators expressed in coordinate space, which have explicit forms as follows:

\begin{eqnarray}
	&&S_{q}^{ab}(x)=i\delta _{ab}\frac{\slashed x}{2\pi ^{2}x^{4}}-\delta _{ab}%
	\frac{m_{q}}{4\pi ^{2}x^{2}}-\delta _{ab}\frac{\langle \overline{q}q\rangle
	}{12}+i\delta _{ab}\frac{\slashed xm_{q}\langle \overline{q}q\rangle }{48}%
	-\delta _{ab}\frac{x^{2}}{192}\langle \overline{q}g_{s}\sigma Gq\rangle
	\notag \\
	&&+i\delta _{ab}\frac{x^{2}\slashed xm_{q}}{1152}\langle \overline{q}%
	g_{s}\sigma Gq\rangle -i\frac{g_{s}G_{ab}^{\alpha \beta }}{32\pi ^{2}x^{2}}%
	\left[ \slashed x{\sigma _{\alpha \beta }+\sigma _{\alpha \beta }}\slashed x%
	\right] -i\delta _{ab}\frac{x^{2}\slashed xg_{s}^{2}\langle \overline{q}%
		q\rangle ^{2}}{7776}  \notag \\
	&&-\delta _{ab}\frac{x^{4}\langle \overline{q}q\rangle \langle
		g^2_{s}G^{2}\rangle }{27648}+\cdots ,  \label{eq:A1}
\end{eqnarray}%
and
\begin{eqnarray}
	&&S_{Q}^{ab}(x)=i\int \frac{d^{4}k}{(2\pi )^{4}}e^{-ikx}\Bigg \{\frac{\delta
		_{ab}\left( {\slashed k}+m_{Q}\right) }{k^{2}-m_{Q}^{2}}-\frac{%
		g_{s}G_{ab}^{\alpha \beta }}{4}\frac{\sigma _{\alpha \beta }\left( {\slashed %
			k}+m_{Q}\right) +\left( {\slashed k}+m_{Q}\right) \sigma _{\alpha \beta }}{%
		(k^{2}-m_{Q}^{2})^{2}}  \notag  \label{eq:A2} \\
	&&+\frac{g_{s}^{2}G^{2}}{12}\delta _{ab}m_{Q}\frac{k^{2}+m_{Q}{\slashed k}}{%
		(k^{2}-m_{Q}^{2})^{4}}+\frac{g_{s}^{3}G^{3}}{48}\delta _{ab}\frac{\left( {%
			\slashed k}+m_{Q}\right) }{(k^{2}-m_{Q}^{2})^{6}}\left[ {\slashed k}\left(
	k^{2}-3m_{Q}^{2}\right) +2m_{Q}\left( 2k^{2}-m_{Q}^{2}\right) \right] \left(
	{\slashed k}+m_{Q}\right) +\cdots \Bigg \},  \notag \\
	&&
\end{eqnarray}
where $G_{\mu\nu}$ is the gluon field strength tensor,  $G_{ab}^{\alpha\beta}=G_A^{\alpha\beta}t^A_{ab}$, $t^A=\lambda^A/2$, $G^{2}=G_{\alpha \beta }^{A}G_{A}^{\alpha \beta }$ and $G^{3}=f^{ABC}G_{\alpha
	\beta }^{A}G^{B\beta \delta }G_{\delta }^{C\alpha }$. $\lambda
	^{A}$ are the Gell-Mann matrices and $f^{ABC}$ are the color structure constants of
	the $SU_{c}(3)$ group where the indices A, B and C take the values from 1 to 8. In the QCD side, the correlation function is evaluated in the deep Euclidean region by using the OPE. After applying the Wick's theorem and conducting all contractions of the quark fields for the symmetric part, we obtain the subsequent expression in relation to the heavy and light quarks propagators:
\begin{align} \label{shekari}
	\Pi_{QCD}^{S}(q)&=i A \epsilon_{a b c} \epsilon_{a^{\prime} b^{\prime} c^{\prime}} \int d^{4} x e^{i q x} \Bigg\{-\gamma_{5} S_{Q}^{c b^{\prime}} S_{q}^{\prime b a^{\prime}} S_{Q^{\prime}}^{a c^{\prime}} \gamma_{5}-\gamma_{5} S_{Q^{\prime}}^{c b^{\prime}} S_{q}^{\prime b a^{\prime}} S_{Q}^{a c^{\prime}} \gamma_{5} +\gamma_{5} S_{Q^{\prime}}^{c c^{\prime}} \gamma_{5} Tr\Big[S_{Q}^{a b^{\prime}} S_{q}^{\prime b a^{\prime}}\Big]\nonumber\\
	&+\gamma_{5} S_{Q}^{c c^{\prime}} \gamma_{5} Tr\Big[S_{Q^{\prime}}^{a b^{\prime}} S_{q}^{\prime b a^{\prime}}\Big] +t \Bigg(-\gamma_{5} S_{Q}^{c b^{\prime}} \gamma_{5} S_{q}^{\prime b a^{\prime}} S_{Q^{\prime}}^{a c^{\prime}}-\gamma_{5} S_{Q^{\prime}}^{c b^{\prime}} \gamma_{5} S_{q}^{\prime b a^{\prime}} S_{Q}^{a c^{\prime}}-S_{Q}^{c b^{\prime}} S_{q}^{\prime b a^{\prime}} \gamma_{5} S_{Q^{\prime}}^{a c^{\prime}} \gamma_{5}\nonumber\\
	&-S_{Q^{\prime}}^{c b^{\prime}} S_{q}^{\prime b a^{\prime}} \gamma_{5} S_{Q}^{a c^{\prime}} \gamma_{5} +\gamma_{5} S_{Q^{\prime}}^{c c^{\prime}}Tr\Big[S_{Q}^{a b^{\prime}} \gamma_{5} S_{q}^{\prime b a^{\prime}}\Big]+S_{Q^{\prime}}^{c c^{\prime}} \gamma_{5} Tr\Big[S_{Q}^{a b^{\prime}} S_{q}^{\prime b a^{\prime}} \gamma_{5}\Big]+\gamma_{5} S_{Q}^{c c^{\prime}} Tr\Big[S_{Q^{\prime}}^{a b^{\prime}} \gamma_{5} S_{q}^{\prime b a^{\prime}}\Big]\nonumber\\
	&+S_{Q}^{c c^{\prime}} \gamma_{5} Tr\Big[S_{Q^{\prime}}^{a b^{\prime}} S_{q}^{\prime b a^{\prime}} \gamma_{5}\Big]\Bigg) +t^{2}\Bigg(-S_{Q}^{c b^{\prime}} \gamma_{5} S_{q}^{\prime b a^{\prime}} \gamma_{5} S_{Q^{\prime}}^{a c^{\prime}}-S_{Q^{\prime}}^{c b^{\prime}} \gamma_{5} S_{q}^{\prime b a^{\prime}} \gamma_{5} S_{Q}^{a c^{\prime}}+S_{Q^{\prime}}^{c c^{\prime}} Tr\Big[S_{q}^{b a^{\prime}} \gamma_{5} S_{Q}^{\prime a b^{\prime}} \gamma_{5}\Big]\nonumber\\
	&+S_{Q}^{c c^{\prime}} Tr\Big[S_{q}^{b a^{\prime}} \gamma_{5} S_{Q^{\prime}}^{\prime a b^{\prime}} \gamma_{5}\Big]\Bigg)\Bigg\},
\end{align}
where $S' = CS^T C$. If $ Q'=Q$ , the value of constant A becomes 1 and when we have $ Q' \neq Q$, this value becomes $\frac{1}{2}$. We also have the antisymmetric part as
\begin{align}\label{e3113}
	\Pi_{QCD}^{A}(q)&=\frac{i}{6} \epsilon_{a b c} \epsilon_{a^{\prime} b^{\prime} c^{\prime}} \int d^{4} x e^{i q x}\Bigg\{2 \gamma_{5} S_{Q}^{c b^{\prime}} S_{Q^{\prime}}^{\prime a a^{\prime}} S_{q}^{b c^{\prime}} \gamma_{5}+\gamma_{5} S_{Q}^{c b^{\prime}} S_{q}^{\prime b a^{\prime}} S_{Q^{\prime}}^{a c^{\prime}} \gamma_{5}-2 \gamma_{5} S_{Q^{\prime}}^{c a^{\prime}} S_{Q}^{\prime a b^{\prime}} S_{q}^{b c^{\prime}} \gamma_{5}+\gamma_{5} S_{Q^{\prime}}^{c b^{\prime}} S_{q}^{\prime b a^{\prime}} S_{Q}^{a c^{\prime}} \gamma_{5}\nonumber\\
	&-2 \gamma_{5} S_{q}^{c a^{\prime}} S_{Q}^{\prime a b^{\prime}} S_{Q^{\prime}}^{b c^{\prime}} \gamma_{5}+2 \gamma_{5} S_{q}^{c a^{\prime}} S_{Q^{\prime}}^{\prime b b^{\prime}} S_{Q}^{a c^{\prime}} \gamma_{5}+4 \gamma_{5} S_{q}^{c c^{\prime}} \gamma_{5} Tr \Big[S_{Q}^{a b^{\prime}} S_{Q^{\prime}}^{b a^{\prime}}\Big]+\gamma_{5} S_{Q^{\prime}}^{c c^{\prime}} \gamma_{5} Tr\Big[S_{Q}^{a b^{\prime}} S_{q}^{\prime b a^{\prime}}\Big]+\nonumber\\
	&\gamma_{5} S_{Q}^{c c^{\prime}} \gamma_{5}  Tr\Big[S_{Q^{\prime}}^{a b^{\prime}} S_{q}^{\prime b a^{\prime}}\Big]+t\Bigg(2 \gamma_{5} S_{Q}^{c b^{\prime}} \gamma_{5} S_{Q^{\prime}}^{\prime a a^{\prime}} S_{q}^{b c^{\prime}}+\gamma_{5} S_{Q}^{c b^{\prime}} \gamma_{5} S_{q}^{\prime b a^{\prime}} S_{Q^{\prime}}^{a c^{\prime}}-2 \gamma_{5} S_{Q^{\prime}}^{c a^{\prime}} \gamma_{5} S_{Q}^{\prime a b^{\prime}} S_{q}^{b c^{\prime}}+\gamma_{5} S_{Q^{\prime}}^{c b^{\prime}} \gamma_{5} S_{q}^{\prime b a^{\prime}} S_{Q}^{a c^{\prime}}\nonumber\\
	&-2 \gamma_{5} S_{q}^{c a^{\prime}} \gamma_{5} S_{Q}^{\prime a b^{\prime}} S_{Q^{\prime}}^{b c^{\prime}}+2 \gamma_{5} S_{q}^{c a^{\prime}} \gamma_{5} S_{Q^{\prime}} b^{b^{\prime}} S_{Q}^{a c^{\prime}}+2 S_{Q}^{c b^{\prime}} S_{Q^{\prime}}^{\prime a a^{\prime}} \gamma_{5} S_{q}^{b c^{\prime}} \gamma_{5}+S_{Q}^{c b^{\prime}} S_{q}^{\prime b a^{\prime}} \gamma_{5} S_{Q^{\prime}}^{a c^{\prime}} \gamma_{5}-2 S_{Q^{\prime}}^{c a^{\prime}} S_{Q}^{\prime a b^{\prime}} \gamma_{5} S_{q}^{b c^{\prime}} \gamma_{5}\nonumber\\
	&+S_{Q^{\prime}}^{c b^{\prime}} S_{q}^{\prime b a^{\prime}} \gamma_{5} S_{Q}^{a c^{\prime}} \gamma_{5}-2 S_{q}^{c a^{\prime}} S_{Q}^{\prime a b^{\prime}} \gamma_{5} S_{Q^{\prime}}^{b c^{\prime}} \gamma_{5}+2 S_{q}^{c a^{\prime}} S_{Q^{\prime}}^{\prime b b^{\prime}} \gamma_{5} S_{Q}^{a c^{\prime}} \gamma_{5}+4 \gamma_{5} S_{q}^{c c^{\prime}} Tr\Big[S_{Q}^{a b^{\prime}} \gamma_{5} S_{Q^{\prime}}^{\prime b a^{\prime}}\Big]\nonumber\\
	&+4 S_{q}^{c c^{\prime}} \gamma_{5}Tr\Big[S_{Q}^{a b^{\prime}} S_{Q^{\prime}}^{\prime b a^{\prime}} \gamma_{5}\Big]+\gamma_{5} S_{Q^{\prime}}^{c c^{\prime}} Tr\Big[S_{Q}^{a b^{\prime}} \gamma_{5} S_{q}^{\prime b a^{\prime}}\Big]+S_{Q^{\prime}}^{c c^{\prime}} \gamma_{5} Tr\Big[S_{Q}^{a b^{\prime}} S_{q}^{\prime b a^{\prime}} \gamma_{5}\Big]+\gamma_{5} S_{Q}^{c c^{\prime}} Tr\Big[S_{Q^{\prime}}^{a b^{\prime}} \gamma_{5} S_{q}^{\prime b a^{\prime}}\Big]\nonumber\\
	&+S_{Q}^{c c^{\prime}} \gamma_{5} Tr\Big[S_{Q^{\prime}}^{a b^{\prime}} S_{q}^{\prime b a^{\prime}} \gamma_{5}\Big]\Bigg)+t^{2}\Bigg(2 S_{Q}^{c b^{\prime}} \gamma_{5} S_{Q^{\prime}}^{a a a^{\prime}} \gamma_{5} S_{q}^{b c^{\prime}}+S_{Q}^{c b^{\prime}} \gamma_{5} S_{q}^{\prime b a^{\prime}} \gamma_{5} S_{Q^{\prime}}^{a c^{\prime}}-2 S_{Q^{\prime}}^{c a^{\prime}} \gamma_{5} S_{Q}^{\prime a b^{\prime}} \gamma_{5} S_{q}^{b c^{\prime}}\nonumber\\
	&+S_{Q^{\prime}}^{c b^{\prime}} \gamma_{5} S_{q}^{\prime b a^{\prime}} \gamma_{5} S_{Q}^{a c^{\prime}}-2 S_{q}^{c a^{\prime}} \gamma_{5} S_{Q}^{\prime a b^{\prime}} \gamma_{5} S_{Q^{\prime}}^{b c^{\prime}}+2 S_{q}^{c a^{\prime}} \gamma_{5} S_{Q^{\prime}}^{\prime b b^{\prime}} \gamma_{5} S_{Q}^{a c^{\prime}}+4 S_{q}^{c c^{\prime}} Tr\Big[S_{Q^{\prime}}^{b a^{\prime}} \gamma_{5} S_{Q}^{\prime a b^{\prime}} \gamma_{5}\Big]\nonumber\\
	&+S_{Q^{\prime}}^{c c^{\prime}} Tr\Big[S_{q}^{b a^{\prime}} \gamma_{5} S_{Q}^{\prime a b^{\prime}} \gamma_{5}\Big]+S_{Q}^{c c^{\prime}} Tr\Big[S_{q}^{b a^{\prime}} \gamma_{5} S_{Q^{\prime}}^{\prime a b^{\prime}} \gamma_{5}\Big]\Bigg)\Bigg\}.
\end{align}

In order to obtain the result for the QCD side of the sum rule, the above mentioned propagators are first used and the calculations are followed by Fourier and Borel transformations. Finally, the process of continuum subtraction is carried out, assisted by the quark-hadron duality assumption. The result of the QCD side is presented in the form
\begin{equation}
	\tilde\Pi_{1(2)}^{S(A)}(q)=\int_{(m_{Q}+m_{Q\prime})^2}^{s_0} ds e^{-s/M^2} \rho_{1(2)}^{S(A)}(s) + \Gamma _{1(2)}^{S(A)}(M^2),
\end{equation}
where subindex 1(2) corresponds to the coefficient of the Lorentz structure $\not\!q (I)$, $s_0$ demonstrates the continuum threshold and $\rho _{1(2)}^{S(A)}(s)$ is the spectral density that is achieved by taking the
imaginary part of the correlation function, $\rho_{1(2)}^{S(A)}(s)=\frac{1}{\pi} \mathrm{Im} [\tilde\Pi _{1(2)}^{S(A)}(q)]$. In the appendix, we present the expressions of the spectral densities obtained after calculations up to 5 nonperturbative operators. $\Gamma_{1(2)}^{S(A)}(M^2)  $ is a function that presents the calculations of the nonperturbative part from the mass dimension 6 to 10. The $\Gamma_{1(2)}^{S(A)}(M^2)  $ is a very lengthy function, so we do not present its explicit form in the present study.

The results obtained from performing computations on both sides are compared using dispersion relations that take into account the coefficients of the same Lorentz structures that are $\not\!q$ and $I$. This yields the QCD sum rules for the relevant quantities, which can be expressed as:
\begin{eqnarray}
	\lambda^2 e^{-\frac{m^2}{M^2}}+\tilde{\lambda}^2e^{-\frac{\tilde{m}^2}{M^2}}+\lambda'^2e^{-\frac{m'{}^2}{M^2}}=\tilde{\Pi}^{S(A)}_{\not\!q}(s_0,M^2),
	\label{Eq:cor:match1}
\end{eqnarray}
and
\begin{eqnarray}
	\lambda^2 m e^{-\frac{m^2}{M^2}}-\tilde{\lambda}^2\tilde{m}e^{-\frac{\tilde{m}^2}{M^2}}+\lambda'^2m'e^{-\frac{m'{}^2}{M^2}}=\tilde{\Pi}^{S(A)}_{I}(s_0,M^2).
	\label{Eq:cor:match2}
\end{eqnarray}
In the following, we show how one can derive QCD sum rules for the masses and residues of the ground  and excited states for the $\not\!q$ structure. To achieve this, one follows a three-step procedure. First, the mass and residue for only the ground state are derived.  Hence, we follow the ground state+continuum scheme, in which we treat the second and third terms on the left-hand side of  Eq.~(\ref{Eq:cor:match1})  as components of the continuum. By this way, we choose only the ground state contribution by adjusting the continuum threshold such that only the ground state is produced in the physical side. Considering only the first term on the left-hand side of  Eq.~(\ref{Eq:cor:match1}),  we obtain the following sum rule for the mass

\begin{eqnarray}
	m^2=\frac{\frac{d}{d(-\frac{1}{M^2})}\tilde{\Pi}^{S(A)}_{\not\!q}(s_0,M^2)}{\tilde{\Pi}^{S(A)}_{\not\!q}(s_0,M^2)},
	\label{Eq:mass:Groundstate}
\end{eqnarray}
where we applied the derivative with respect to $-\frac{1}{M^2}  $ to both sides and divided the resultant equation by the original one.
The residue is obtained as
\begin{eqnarray}
	\lambda^2=e^{\frac{m^2}{M^2}}\tilde{\Pi}^{S(A)}_{\not\!q}(s_0,M^2).
	\label{Eq:Groundstate}
\end{eqnarray}
In the next step,  having calculated the mass and residue of the  ground state, we increase the threshold parameter to bring the first orbital excited state, $1P$,  to the physical side. In this step we consider the first two terms on the left-hand side of Eq.~(\ref{Eq:cor:match1}), representing  the ground  and  $1P$ states, respectively. By this way, we  considered  the third term representing  the $2S$ state inside the continuum and applied the ground state + first orbitally excited state +  continuum scheme. Now,  considering the results of the mass and residue for the ground state, we have two new unknowns, which are the mass and residue of  the $1P$ state. We can obtain them from the similar way as the ground state. Finally, we follow the ground state + first orbitally excited state + first radially excited state + continuum scheme to achieve the mass and residue of the radially excited, $2S$, state by increasing again the value of the continuum threshold and using the parameters of the ground and $1P$ states as inputs. We will come back to these procedures in next section considering the related numerical values.

\section{Numerical Analysis }
In this section, we present the numerical analysis of our results for the masses and residues of the doubly heavy baryons in their ground, first orbitally and first radially excited states. We consider some inputs that are shown in table~\ref{tab:Parameter}. In the calculations, the masses of $u$ and $d$ quarks are assumed as zero, but we keep the $s$ quark mass. As seen from table~\ref{tab:Parameter}, we display the values of condensates up to six mass dimensions, but we perform calculations up to ten mass dimensions. The  nonperturbative operators with mass dimensions greater than six (seven to ten) are written  in terms of low-dimensional operators using the QCD factorization hypothesis. This is done when calculating the expressions of the heavy and light quarks propagators presented in the previous section. When the propagators are multiplied through Eqs. (\ref{shekari}) and (\ref{e3113}) the higher dimensional operators appear as the multiplications of the operators with lower dimensions, whose values are obtained via the numerical values presented in  table~\ref{tab:Parameter}.

\begin{figure}[h!]
	\begin{center}
		\includegraphics[totalheight=6cm,width=8cm]{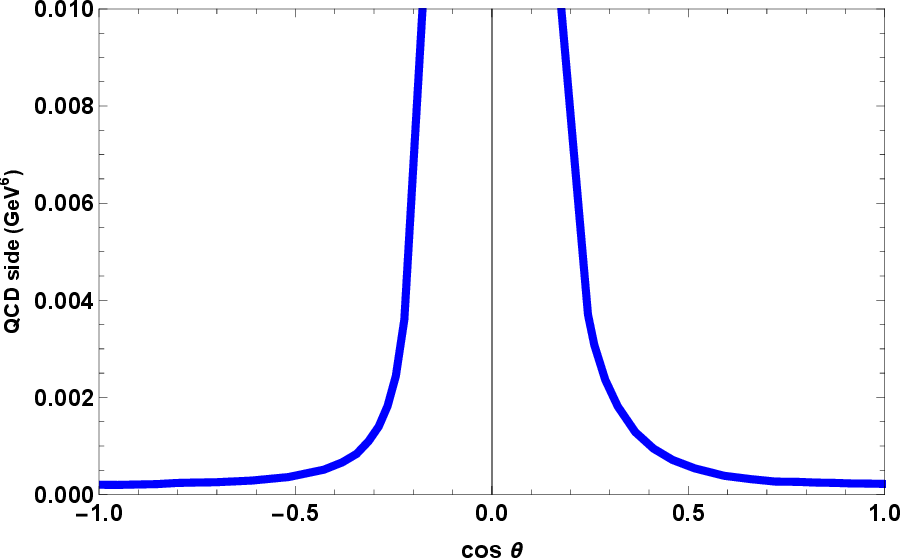}
	\end{center}
	\caption{  $ \tilde{\Pi}^{\mathrm{QCD}}_{\not\!q} $ as a function of $ \cos\theta $ at the central values of $M^2$ and $s_0$.}
	\label{fig:MassT}
\end{figure}

\begin{figure}[h!]
	\begin{center}
		\includegraphics[totalheight=6cm,width=8cm]{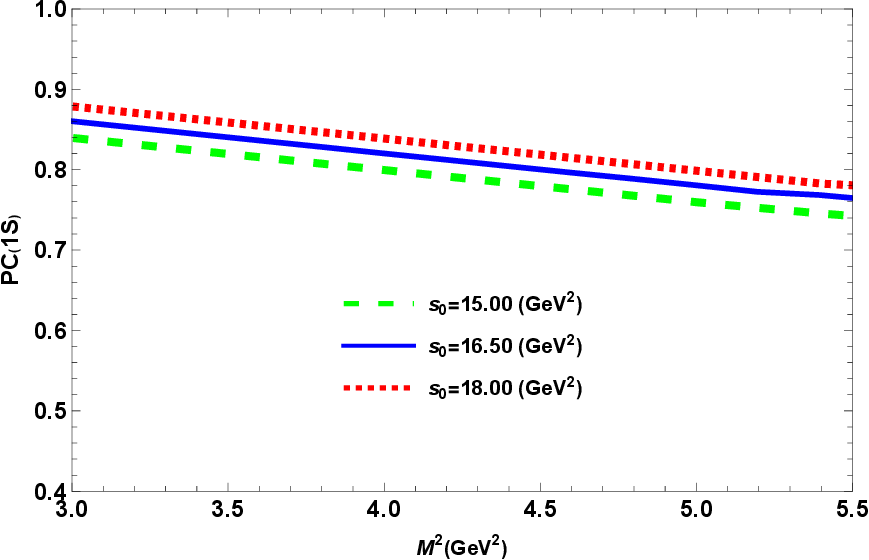}
	\end{center}
	\caption{ Pole contribution of $	\Xi_{cc}$ as a function of the Borel parameter $M^2$ at fixed values of $s_0$.}
	\label{fig:pcchicc}
\end{figure}

\begin{table}[tbp]
	\begin{tabular}{|c|c|}
		\hline
		Parameters & Values \\ \hline\hline
		$m_{c}$                                     & $1.27\pm 0.02~\mathrm{GeV}$ \cite{Zyla:2020zbs}\\
		$m_{b}$                                     & $4.18^{+0.03}_{-0.02}~\mathrm{GeV}$ \cite{Zyla:2020zbs}\\
		$m_{s}$                                   & $93^{+11}_{-5}~\mathrm{MeV}$ \cite{Zyla:2020zbs}\\
		$\langle \bar{q}q \rangle $    & $(-0.24\pm 0.01)^3$ $\mathrm{GeV}^3$ \cite{Belyaev:1982sa}  \\
		$\langle \bar{s}s \rangle $               & $0.8\langle \bar{q}q \rangle$ \cite{Belyaev:1982sa} \\
		$m_{0}^2 $                                & $(0.8\pm0.1)$ $\mathrm{GeV}^2$ \cite{Belyaev:1982sa}\\
		$\langle \overline{q}g_s\sigma Gq\rangle$ & $m_{0}^2\langle \bar{q}q \rangle$ \\
		$\langle g_s^2 G^2 \rangle $              & $4\pi^2 (0.012\pm0.004)$ $~\mathrm{GeV}
		^4 $\cite{Belyaev:1982cd}\\
		$\langle \frac{\alpha_s}{\pi} G^2 \rangle $ & $(0.012\pm0.004)$ $~\mathrm{GeV}^4 $\cite{Belyaev:1982cd}\\
		$\langle g_s^3 G^3 \rangle $                & $ (0.57\pm0.29)$ $~\mathrm{GeV}^6 $\cite{Narison:2015nxh}\\
		
		\hline\hline
	\end{tabular}%
	\caption{Some input parameters entering the calculations.}
	\label{tab:Parameter}
\end{table}
In addition to the parameters listed in table~\ref{tab:Parameter}, there are three extra parameters, namely Borel parameter $M^2$, threshold parameter $s_0$ and arbitrary mixing parameter $t$. They are obtained from the analysis of results based on the standard criteria of the QCD sum rule method. These criteria include weak dependence of the results on auxiliary parameters, pole dominance and convergence of the OPE. We determine the working regions of $t$ from the analysis by considering a parametric plot of the results as functions of $\cos\theta$, where $t=\tan\theta$. The QCD side of calculations in terms of $ \cos\theta $ is shown in Fig.~\ref{fig:MassT}, which is valid for all states of the doubly heavy baryons.
We select the regions that show least variations with respect to the changes in $\cos\theta$, which read
\begin{eqnarray}
	-1\leq\cos\theta\leq -0.5 ~~~~~\mbox{and} ~~~~~~0.5\leq \cos\theta\leq 1. 
\end{eqnarray}
 Additionally, we determine the working region of Borel mass through the following criteria. The upper bound for $M^2$, is set by ensuring that the pole contribution should be larger than the continuum and higher states contributions. To meet this requirement, we stipulate that the ratio $ PC $,

\begin{eqnarray}
	PC=\frac{\Pi(s_0,M^2,t)}{\Pi(\infty,M^2,t)},
\end{eqnarray}
should surpass $0.50$. In Fig.~\ref{fig:pcchicc}, we can see the satisfaction of this condition for pole contribution of $\Xi_{cc}$ as a function of the Borel parameter $M^2$ at fixed values of $s_0$.
The lower bound for $M^2$ is set based on the condition that the OPE series must be convergent, which means that the perturbative contribution should be greater than the nonperturbative one and the higher the dimension of the nonperturbative operator, the lower its contribution. For this purpose, we introduce the ratio:
\begin{equation}
	R(M^{2})=\frac{\Pi ^{\mathrm{DimN}}(s_0,M^2,t)}{\Pi (s_0,M^2,t)},
	\label{eq:Convergence}
\end{equation}

Where $\mathrm{DimN}$ refers to the last three dimensions, i.e., $\mathrm{DimN=Dim(8+9+10)}$. To determine the minimum of the Borel parameter, the ratio should not exceed $0.05$ of the total contribution. As the final parameter, the continuum threshold values $s_0$, which also depend on the energy of the next first excited state, are chosen in such a way that the integrals do not receive a contribution from the excited states during the ground state calculations. Table~\ref{tab:results} presents the working intervals for the Borel parameter $M^2$ and threshold parameter $s_0$ for all the channels.

\begin{figure}[h!]
	\begin{center}
		\includegraphics[totalheight=6cm,width=8cm]{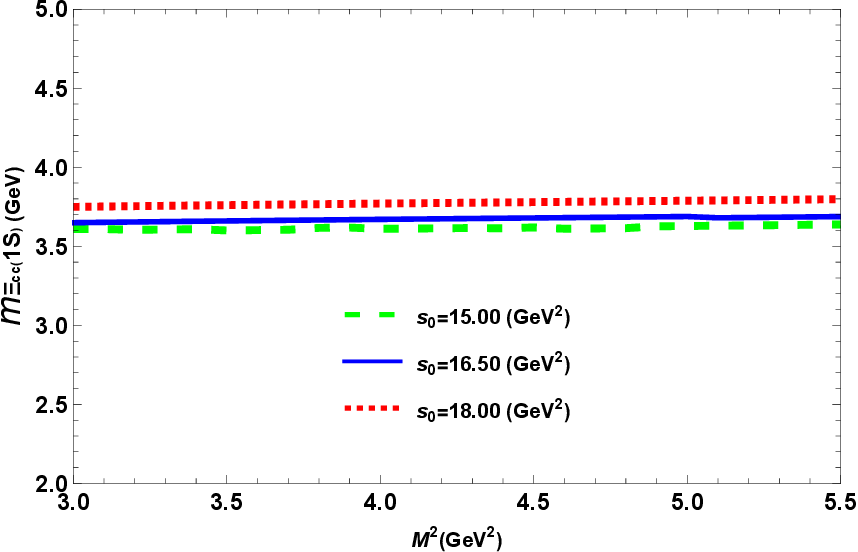}
		\includegraphics[totalheight=6cm,width=8cm]{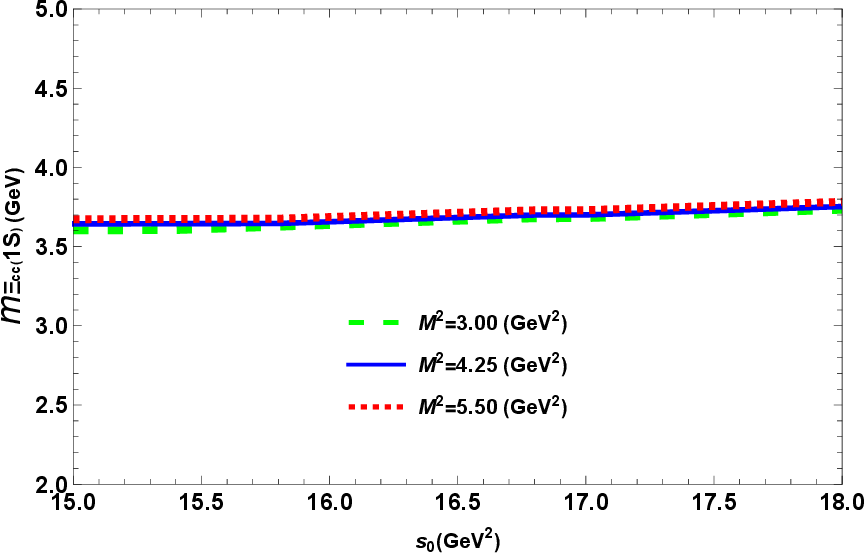}
	\end{center}
	\caption{\textbf{Left:} the variation of the mass of $	\Xi_{cc}$ at $1S $ state as a function of the Borel parameter $M^2$ and at the various amounts of the parameter $s_0$. \textbf{Right:} the variation of the mass of $	\Xi_{cc}$ at $1S $ state as a function of the threshold parameter $s_0$ and at the various amounts of the parameter $M^2$.}
	\label{gr:g1}
\end{figure}

\begin{figure}[h!]
	\begin{center}
		\includegraphics[totalheight=6cm,width=8cm]{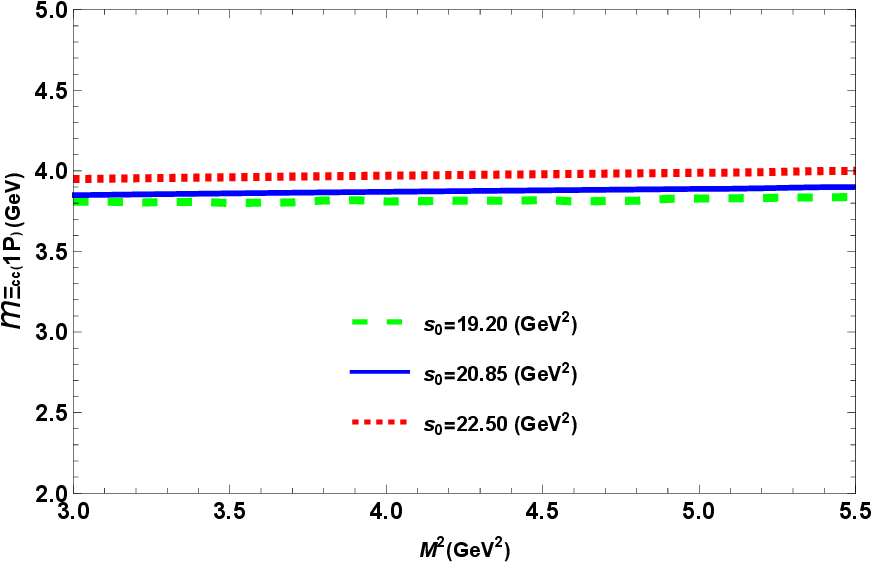}
		\includegraphics[totalheight=6cm,width=8cm]{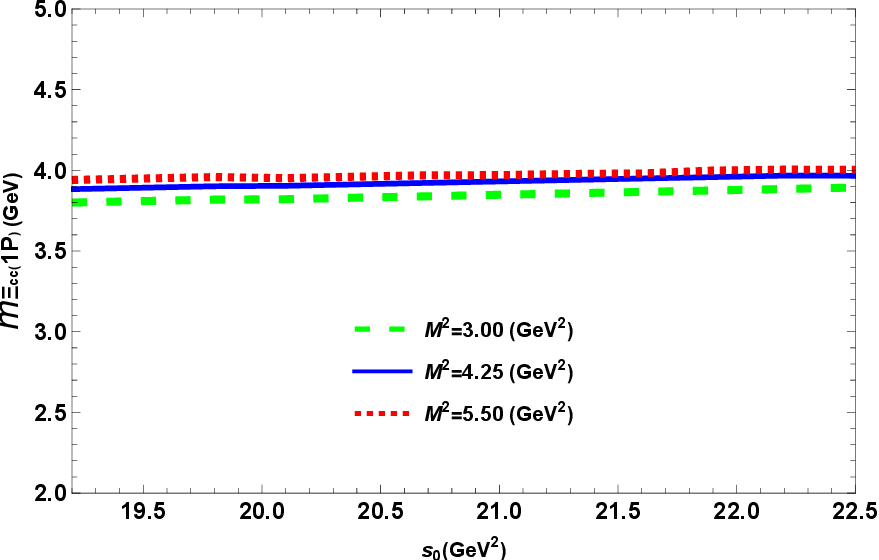}
	\end{center}
	\caption{\textbf{Left:} the variation of the mass of $	\Xi_{cc}$ at $1P$ state as a function of the Borel parameter $M^2$ and at the various amounts of the parameter $s_0$. \textbf{Right:} the variation of the mass of $	\Xi_{cc}$ at $1P $ state as a function of the threshold parameter $s_0$ and at the various amounts of the parameter $M^2$.}
	\label{gr:g2}
\end{figure}

\begin{figure}[h!]
	\begin{center}
		\includegraphics[totalheight=6cm,width=8cm]{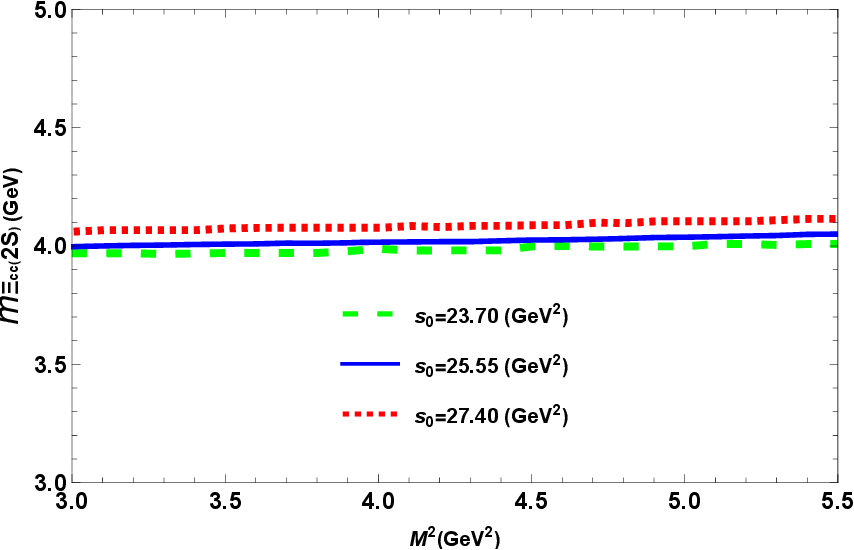}
		\includegraphics[totalheight=6cm,width=8cm]{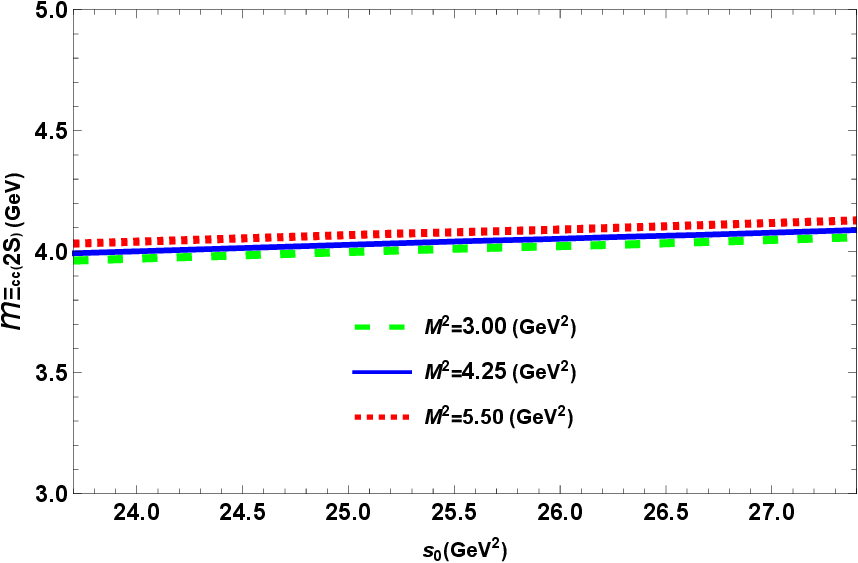}
	\end{center}
	\caption{\textbf{Left:} the variation of the mass of $	\Xi_{cc}$ at $2S $ state as a function of the Borel parameter $M^2$ and at the various amounts of the parameter $s_0$. \textbf{Right:} the variation of the mass of $	\Xi_{cc}$ at $2S $ state as a function of the threshold parameter $s_0$ and at the various amounts of the parameter $M^2$.}
	\label{gr:g3}
\end{figure}

Based on our analysis, we have observed that there is only a weak dependence between the physical quantities and the auxiliary parameters in the given windows for $M^2$ and $s_0$. To portray how the masses are affected by the auxiliary parameters, we include Figs.~\ref{gr:g1}, \ref{gr:g2}, and \ref{gr:g3}. We draw these figures at the average value of the arbitrary mixing parameter, i.e.,  $t=0.75$. The results demonstrate a high level of stability in relation to the Borel parameter and continuum threshold within their respective working regions. It should be noted that our results for all baryons show least dependence on auxiliary parameters. For this reason, we include only the graphs for the $\Xi_{cc}$ state. The principal sources of uncertainties in our numerical outcomes are due to the uncertainties with respect to the auxiliary parameters and errors of other input parameters.

 After determination of the range of auxiliary parameters, we present the mass and residue of the doubly heavy baryons with spin-$\frac{1}{2}$, obtained through numerical analyses in table~\ref{tab:results}.

 \begin{table}[]
 	\begin{tabular}{|c|c|c|c|c|c|}
 		\hline
 		Baryon  & State &$M^2~(\mathrm{GeV^2})$&$s_0~(\mathrm{GeV^2})$  & $\mathrm {Mass}~(\mathrm{GeV})$ & $\mathrm {Residue}~(\mathrm{GeV^3})$ \\ \hline\hline
 		&$	\Xi_{cc}(\frac{1}{2}^+)(1S)$ &$3.00-5.50$& $3.87^2-4.24^2$& $3.69  \pm0.10$ & $0.16  \pm0.04$  \\ \cline{2-6} 
 		$	\Xi_{cc}$   &$\Xi_{cc}(\frac{1}{2}^-)(1P)$ &$3.00-5.50$& $4.37^2-4.74^2$& $3.91 _{-0.11} ^{+0.09}$ & $0.18  \pm0.03$  \\ \cline{2-6} 
 		&$\Xi_{cc}(\frac{1}{2}^+)(2S)$ &$3.00-5.50$& $4.87^2-5.24^2$& $4.04  \pm0.08$ & $0.19  \pm0.03$ \\ \hline\hline
 		&$\Xi_{bc}(\frac{1}{2}^+)(1S)$ &$6.00-9.00$& $6.92^2-7.21^2$& $6.73 _{-0.13} ^{+0.14}$ & $0.29  \pm0.06$ \\ \cline{2-6} 
 		$\Xi_{bc}$     &$\Xi_{bc}(\frac{1}{2}^-)(1P)$ &$6.00-9.00$& $7.42^2-7.71^2$& $6.94  \pm0.13$ & $0.32 _{-0.04} ^{+0.06}$ \\ \cline{2-6} 
 		&$\Xi_{bc}(\frac{1}{2}^+)(2S)$ &$6.00-9.00$& $7.92^2-8.21^2$& $7.12  \pm0.14$ & $0.33  \pm0.06$ \\ \hline\hline
 		&$	\Xi_{bb}(\frac{1}{2}^+)(1S)$ &$10.00-15.00$& $10.58^2-10.86^2$& $9.97  \pm0.19$ & $0.45_{-0.08} ^{+0.09}$  \\ \cline{2-6} 
 		$	\Xi_{bb}$   &$\Xi_{bb}(\frac{1}{2}^-)(1P)$ &$10.00-15.00$& $11.08^2-11.36^2$& $10.25  \pm0.18$ & $0.60 _{-0.08} ^{+0.09}$  \\ \cline{2-6} 
 		&$\Xi_{bb}(\frac{1}{2}^+)(2S)$ &$10.00-15.00$& $11.58^2-11.86^2$& $10.33 _{-0.19} ^{+0.18} $ & $0.70  _{-0.10} ^{+0.09}$ \\ \hline\hline
 		&$\Omega_{cc}(\frac{1}{2}^+)(1S)$ &$3.00-5.50$& $3.89^2-4.26^2$& $3.70  \pm0.09$ & $0.17  \pm0.04$ \\ \cline{2-6} 
 		$\Omega_{cc}$     &$\Omega_{cc}(\frac{1}{2}^-)(1P)$ &$3.00-5.50$& $4.39^2-4.76^2$& $3.93_{-0.09} ^{+010}$ & $0.19_{-0.04} ^{+0.03}$ \\ \cline{2-6} 
 		&$\Omega_{cc}(\frac{1}{2}^+)(2S)$ &$3.00-5.50$& $4.89^2-5.26^2$& $4.07_{-0.09} ^{+0.08}$ & $0.20  _{-0.03} ^{+0.04}$ \\ \hline\hline
 		&$\Omega_{bc}(\frac{1}{2}^+)(1S)$ &$6.00-9.00$&$6.92^2-7.21^2$& $6.77_{-0.12} ^{+0.13}$ & $0.30  \pm0.05$ \\ \cline{2-6} 
 		$\Omega_{bc}$     &$\Omega_{bc}(\frac{1}{2}^-)(1P)$ &$6.00-9.00$& $7.42^2-7.71^2$& $7.07  \pm0.12$ & $0.33 _{-0.06} ^{+0.05}$ \\ \cline{2-6} 
 		&$\Omega_{bc}(\frac{1}{2}^+)(2S)$ &$6.00-9.00$& $7.92^2-8.21^2$& $7.20  \pm0.13$ & $0.35  \pm0.06$ \\ \hline\hline
 		&$\Omega_{bb}(\frac{1}{2}^+)(1S)$ &$10.00-15.00$& $10.58^2-10.86^2$& $9.98\pm0.18$ & $0.46_{-0.08} ^{+0.09}$ \\ \cline{2-6} 
 		$\Omega_{bb}$     &$\Omega_{bb}(\frac{1}{2}^-)(1P)$ &$10.00-15.00$& $11.08^2-11.36^2$& $10.31 \pm0.19$ & $0.63  \pm0.09$ \\ \cline{2-6} 
 		&$\Omega_{bb}(\frac{1}{2}^+)(2S)$ &$10.00-15.00$& $11.58^2-11.86^2$& $10.45  \pm0.18$ & $0.75  _{-0.09} ^{+0.08}$ \\ \hline\hline
 		&$\Xi^{\prime}_{bc}(\frac{1}{2}^+)(1S)$ &$6.00-9.00$& $6.92^2-7.21^2$& $6.81  \pm0.11$ & $0.31  \pm0.05$ \\ \cline{2-6} 
 		$\Xi^{\prime}_{bc}$     &$\Xi^{\prime}_{bc}(\frac{1}{2}^-)(1P)$ &$6.00-9.00$& $7.42^2-7.71^2$& $6.99_{-0.12} ^{+0.13}$ & $0.34 _{-0.05} ^{+0.04}$ \\ \cline{2-6} 
 		&$\Xi^{\prime}_{bc}(\frac{1}{2}^+)(2S)$ &$6.00-9.00$& $7.92^2-8.21^2$& $7.15 _{-0.13} ^{+0.14}$ & $0.36  \pm0.06 $ \\ \hline\hline
 		&$\Omega^{\prime}_{bc}(\frac{1}{2}^+)(1S)$ &$6.00-9.00$& $6.92^2-7.21^2$& $6.82  \pm0.12$ & $0.32  \pm0.06$ \\ \cline{2-6} 
 		$\Omega^{\prime}_{bc}$     &$\Omega^{\prime}_{bc}(\frac{1}{2}^-)(1P)$ &$6.00-9.00$&$7.42^2-7.71^2$& $7.03_{-0.12} ^{+0.13}$ & $0.35  \pm0.06$ \\ \cline{2-6} 
 		&$\Omega^{\prime}_{bc}(\frac{1}{2}^+)(2S)$ &$6.00-9.00$& $7.92^2-8.21^2$& $7.24 _{-0.13} ^{+0.14}$ & $0.38  _{-0.07} ^{+0.06}$ \\ \hline\hline
 	\end{tabular}
 	\caption{The auxiliary parameters and the outcomes of the masses and residues for the ground, first orbitally excited and first radially excited states.}
 	\label{tab:results}
 \end{table}

 Our approach involves analyzing the ground state, first orbitally excited state, first radially excited state and continuum, using Eq.~(\ref{Eq:cor:match1}) to progress step-by-step. As previously said, firstly, we derive the mass and residue of the doubly heavy baryons in the ground state using proper threshold parameters based on the ground state+continuum approach. We determine the appropriate range for $s_0$, also listed in table~\ref{tab:results}. Second, we apply the same method to determine the mass and residue of baryons for the ground state, first orbitally excited state and continuum scheme, using suitable threshold parameters. The outcomes of this stage are likewise provided in table~\ref{tab:results}. Finally, we include the radially excited $2S$ state using the ground state, first orbitally excited state, first radially excited state and continuum scheme and specify the appropriate threshold parameter. The results for the $2S$ states can be found in table~\ref{tab:results} as well. It is necessary to mention that the results of two Lorentz structures have roughly the same values. Therefore, we only report the results of $\not\!q$ structure.

As we can see in table~\ref{tab:results}, for $\Xi_{cc}$, the masses of the excited states $1P$ and $2S$ are about 0.22 and 0.35 $\mathrm{GeV}$ more than the mass of the ground state, respectively. These values for other members are as follows: $	\Xi_{bc}$ (0.21 and 0.39 $\mathrm{GeV}$), $	\Xi_{bb}$ (0.28 and 0.36 $\mathrm{GeV}$), $	\Omega_{cc}$ (0.23 and 0.37 $\mathrm{GeV}$), $	\Omega_{bc}$ (0.30 and 0.43 $\mathrm{GeV}$), $	\Omega_{bb}$ (0.33 and 0.47 $\mathrm{GeV}$), $\Xi^{\prime}_{bc}$ (0.18 and 0.34 $\mathrm{GeV}$) and $\Omega^{\prime}_{bc}$ (0.21 and 0.42 $\mathrm{GeV}$).

Calculations of the mass of the doubly heavy baryons in the ground state have been done in different approaches \cite{Zhang:2008rt,Aliev:2012ru,Wang:2010hs,Ebert:2002ig,Narison:2010py}. To provide a comprehensive comparison, table~\ref{tab1} includes predictions for the mass of the doubly heavy baryons obtained from various approaches. We perform these calculations by including the nonperturbative operators of the mass dimensions up to 10 to reduce the errors and increase the accuracy. In the previous work done by QCD sum rule method \cite{Aliev:2012ru}, the calculations were done up to 5 mass dimensions. Our results for the ground states, except for some channels, are overall in good agreements with the predictions of different methods such as relativistic quark model \cite{Ebert:2002ig} and QCD sum rules \cite{Zhang:2008rt,Aliev:2012ru,Wang:2010hs} within the presented uncertainties. Our result on the mass of the $	\Xi_{cc}$ state is consistent within the errors with the experimental data of the LHCb collaboration available only for this channel.

The study of the mass of the doubly heavy baryons for $1P$ and $2S$ excited states have also been done in various methods \cite{Ebert:2002ig,Giannuzzi:2009gh,Aliev:2019lvd,Shah:2016vmd,Shah:2017liu,Wang:2010it,Valcarce:2008dr}. For these excited states, our calculations are reported with higher accuracies as well. As can be seen from tables~\ref{tab3} and ~\ref{tab5}, our results for the $1P$ and $2S$ excited states, except for some channels, are overall in good consistencies with the predictions of various approaches such as relativistic quark model \cite{Ebert:2002ig}, quark model \cite{Yoshida:2015tia}, Salpeter model\cite{Giannuzzi:2009gh}, hypercentral constituent quark model \cite{Shah:2016vmd,Shah:2017liu}, Faddeev method \cite{Valcarce:2008dr} and QCD sum rules \cite{Aliev:2019lvd,Wang:2010it} within the presented errors as well.

 Furthermore, we studied the residues of the ground, first orbitally and first radially excited states of the doubly heavy baryons and reported our outcomes in tables~\ref{tab2},~\ref{tab4} and~\ref{tab6}. For the ground states, using QCD sum rules \cite{Zhang:2008rt,Aliev:2012ru,Wang:2010hs}, researchers have also studied the residue of the doubly heavy baryons. These studies for the $1P$ excited state have been done in QCD sum rules \cite{Aliev:2019lvd,Wang:2010hs}. For the $2S$ excited state, the residue calculation has also been done in Ref. \cite{Aliev:2019lvd}. As mentioned before, our results are more accurate for the residue calculations compared to the previous predictions, since contributions of the nonperturbative operators up to dimension ten are taken into account in the OPE. Although, we see overall good consistencies (there are some differences on some channels) of the results for the masses among different approaches and predictions within the presented uncertainties, overall we see considerable differences in the values of the presented residues for the ground and excited baryons among different studies (there are some consistencies among the results in some channels). The presented uncertainties for the residues are high compared to the masses. This is well understood from the fact that the mass is ratio of two sum rules that leads to killing of the uncertainties in ratio, while the residue is obtained from one sum rule. Our results on the residues can be used to investigate different decay channels of the doubly heavy baryons under study.
 
\begin{table}[tbp]
	\centering
	\begin{tabular}{|c|c|c|c|c|c|c|}
		
		\hline   Baryon& This work & Ref. \cite{Aliev:2012ru} &Ref. \cite{LHCb:2018pcs} & Ref. \cite{Ebert:2002ig} &  Ref. \cite{Zhang:2008rt} &    Ref. \cite{Wang:2010hs}  \\ \hline\hline
		$	\Xi_{cc}$& $3.69  \pm0.10$                         &$3.72  \pm0.20$&$3.62\pm0.0015$&	$3.620 $&$4.26  \pm0.19$&$3.57  \pm0.14$	  \\  
		$	\Xi_{bc}$&$6.73 _{-0.13} ^{+0.14}$                                   &$6.72  \pm0.20$&$-$&	$6.933$&$6.75  \pm0.05$&-	 \\ 
		$	\Xi_{bb}$&$9.97  \pm0.19$                 &$9.96  \pm0.90$&	$-$&$10.202  $ &$9.78\pm 0.07	$&$10.17  \pm0.14$  \\
		$	\Omega_{cc}$&$3.70  \pm0.09$                              & $3.73  \pm0.20$&$-$ &	$3.778 $&$4.25  \pm0.20$&$3.71 \pm0.14$	 \\ 
		$	\Omega_{bc}$&$6.77_{-0.12} ^{+0.13}$   & $6.75  \pm0.30$&$-$&$7.088 $ & $7.02  \pm0.08$	&-	\\
		$	\Omega_{bb}$	&$9.98\pm0.18$              &$9.97  \pm0.90$&$-$&$10.359 $ & $9.85\pm 0.07	$	& $10.32  \pm0.14$\\ 
		$	\Xi^{\prime}_{bc}$&$6.81  \pm0.11$          &$6.79  \pm0.20$&$-$&	$6.963 $ & $6.95  \pm0.08$&-	   \\ 
		$	\Omega^{\prime}_{bc}$&$6.82  \pm0.12$       &$6.80  \pm0.30$&$-$&	$7.116 $ & $7.02  \pm0.08$&-	\\ 
		
		\hline\hline
	\end{tabular}%
	\caption{The masses of  the ground state $1S$ of the doubly heavy baryons in $\mathrm{GeV}$ and comparison of the results with other predictions.}
	\label{tab1}
\end{table}

\begin{table}[tbp]
	\centering
	\begin{tabular}{|c|c|c|c|c|c|c|}
		
		\hline Baryon& This work &  Ref. \cite{Aliev:2019lvd} &  Ref. \cite{Ebert:2002ig} & Ref. \cite{Valcarce:2008dr} &  Ref. \cite{Wang:2010it}  & Ref. \cite{Yoshida:2015tia} \\ \hline\hline
		$	\Xi_{cc}$& $3.91 _{-0.11} ^{+0.09}$     &$4.03  \pm0.20$       &$3.838$&	$3.880 $&$3.77  \pm0.18$	&$3.947$  \\  
		$	\Xi_{bc}$&$6.94  \pm0.13$                       &$7.14  \pm0.11$            &$-$&$-$&	$-$&$-$	 \\ 
		$	\Xi_{bb}$&$10.25  \pm0.18$  &$10.32  \pm0.10$&$10.368$&$10.406  $&$10.38\pm 0.15	$	&$10.476$ \\
		$	\Omega_{cc}$&$3.93_{-0.09} ^{+010}$ &$4.16  \pm0.14$    &$4.002$&	$- $&$3.91  \pm0.14$&$4.086$	  \\ 
		$	\Omega_{bc}$&$7.07  \pm0.12$   &$7.20  \pm0.11$&$-$&$-$&$- $&$-$		\\
		$	\Omega_{bb}$&$10.31 \pm0.19$        &$10.37  \pm0.10$	      &$10.532$&$-$&$10.38\pm 0.15	$	&$10.607$ \\ 
		$	\Xi^{\prime}_{bc}$&$6.99_{-0.12} ^{+0.13}$  &$7.02  \pm0.07$        & $-$&	$- $&$-$&$-$	\\ 
		$	\Omega^{\prime}_{bc}$&$7.03_{-0.12} ^{+0.13}$       &$7.09  \pm0.07$&$-$&	$- $&$-$&$-$	\\ 
		
		\hline\hline
	\end{tabular}%
	\caption{ The masses of $1P$ excitation of the doubly heavy baryons in $\mathrm{GeV}$ and comparison of the results with other predictions.}
	\label{tab3}
\end{table}

\begin{table}[tbp]
	\centering
	\begin{tabular}{|c|c|c|c|c|c|c|c|}
		
		\hline Baryon& This work &  Ref. \cite{Aliev:2019lvd}  &   Ref. \cite{Ebert:2002ig} & Ref. \cite{Shah:2017liu} & Ref. \cite{Giannuzzi:2009gh} &  Ref. \cite{Shah:2016vmd}&Ref. \cite{Yoshida:2015tia}  \\ \hline\hline
		$	\Xi_{cc}$& $4.04  \pm0.08$     &$4.03  \pm0.20$   &$3.910$&$3.920$&	$4.183 $&$-$&$4.079$	\\  
		$	\Xi_{bc}$&$7.12  \pm0.14$              &$7.14  \pm0.11$ &$-$&$7.263$&	$7.495$&$-$&$-$	 \\ 
		$	\Xi_{bb}$&$10.33 _{-0.19} ^{+0.18} 	$   &$10.32  \pm0.10$&$10.441$&$10.609$&$10.751 $&$-	$&$10.571$	 \\
		$	\Omega_{cc}$&$4.07_{-0.09} ^{+0.08} $        &$4.16  \pm0.14$	  &$4.075$&$-$&	$4.268$&$4.041$&$4.227$ \\ 
		$	\Omega_{bc}$&$7.20  \pm0.13$   &$7.20  \pm0.11$&$-$&$-$&$7.559 $&$7.480$	&$-$	 \\
		$	\Omega_{bb}$&$10.45  \pm0.18$              &$10.37  \pm0.10$&$10.610$&$-$&$10.830 $&$10.736	$	&$10.707$	 \\ 
		$	\Xi^{\prime}_{bc}$&$7.15 _{-0.13} ^{+0.14} $          &$7.02  \pm0.07$	& $-$&$-$&	$- $&$-$&$-$  \\ 
		$	\Omega^{\prime}_{bc}$&$7.24 _{-0.13} ^{+0.14} $       &$7.09  \pm0.07$&$-$&$-$&	$- $&$-$&$-$	\\ 
		
		\hline\hline
	\end{tabular}%
	\caption{The masses of $2S$ excitation of the doubly heavy baryons in $\mathrm{GeV}$ and comparison of the results with other predictions.}
	\label{tab5}
\end{table}   

\begin{table}[tbp]
	\centering
	\begin{tabular}{|c|c|c|c|c|}
		\hline  Baryon& This work &   Ref. \cite{Aliev:2012ru}&  Ref. \cite{Zhang:2008rt} & Ref. \cite{Wang:2010hs}   \\ \hline\hline
		$	\Xi_{cc}$&$0.16  \pm0.04$                                    &$0.16  \pm0.02$& $0.042  \pm0.026$&$0.115  \pm0.027$	  \\
		$	\Xi_{bc}$&$0.29  \pm0.06$                                   &$0.28  \pm0.05$&$0.046  \pm0.021$ &-	 \\
		$	\Xi_{bb}$&	$0.45_{-0.08} ^{+0.09}$                              &$0.44  \pm0.08$&$0.067 \pm0.057$&	 $0.252 \pm0.064$ \\
		$	\Omega_{cc}$&$0.17  \pm0.04$                              &$0.18  \pm0.03$&-&$0.138 \pm0.030$	 \\
		$	\Omega_{bc}$&$0.30  \pm0.05$                             &$0.29  \pm0.05$&-&	- \\
		$	\Omega_{bb}$&$0.46_{-0.08} ^{+0.09}$                        &$0.45  \pm0.08$&-&	$0.311\pm0.077$\\
		$	\Xi^{\prime}_{bc}$&$0.31  \pm0.05$                       &$0.30  \pm0.05$	&-& - \\
		$	\Omega^{\prime}_{bc}$&$0.32  \pm0.06$                              &$0.31  \pm0.06$&-&-	\\
		
		\hline\hline
	\end{tabular}%
	\caption{The residues of the ground state $1S$ of the doubly heavy baryons in $\mathrm{GeV^3}$ and comparison of the results with other predictions. }
	\label{tab2}
\end{table}

\begin{table}[tbp]
	\centering
	\begin{tabular}{|c|c|c|c|}
		\hline Baryon& This work\,  & Ref. \cite{Aliev:2019lvd} & Ref. \cite{Wang:2010it}    \\ \hline\hline
		$	\Xi_{cc}$&$0.18  \pm0.03$               &$0.099  \pm0.013$	  &$0.159  \pm0.037$	  \\
		$	\Xi_{bc}$&$0.32 _{-0.04} ^{+0.06}$                           	&$0.242  \pm0.020$     &$-$	 \\
		$	\Xi_{bb}$&$0.60 _{-0.08} ^{+0.09}$                               &$0.576  \pm0.046$&$0.365  \pm0.089$		 \\
		$	\Omega_{cc}$&$0.19_{-0.04} ^{+0.03}  $                      &$0.125  \pm0.015$    &$0.192  \pm0.041$		 \\
		$	\Omega_{bc}$&$0.33 _{-0.06} ^{+0.05}$                   &$0.241  \pm0.021$	   &$-$	 \\
		$	\Omega_{bb}$&$0.63  \pm0.09$                    &$0.620  \pm0.060$ & $0.444  \pm0.101$		\\
		$	\Xi^{\prime}_{bc}$&$0.34 _{-0.05} ^{+0.04} $                      &$0.176  \pm0.084$ & $-$		  \\
		$	\Omega^{\prime}_{bc}$&$0.35  \pm0.06 $                             &$0.174  \pm0.079$  &$-$		\\
		
		\hline\hline
	\end{tabular}%
	\caption{The residues of $1P$ excitation of the doubly heavy baryons in $\mathrm{GeV^3}$ and comparison of the results with other predictions.}
	\label{tab4}
\end{table}

\begin{table}[tbp]
	\centering
	\begin{tabular}{|c|c|c|}
		\hline Baryon& This work\,  & Ref. \cite{Aliev:2019lvd}  \\ \hline\hline
		$	\Xi_{cc}$&$0.19  \pm0.03$                                    &$0.144  \pm0.064$	  \\
		$	\Xi_{bc}$&$0.33  \pm0.06 $                                  &$0.306  \pm0.090$	 \\
		$	\Xi_{bb}$&$0.70  _{-0.10} ^{+0.09} $                               & $0.764  \pm0.114$	\\
		$	\Omega_{cc}$&$0.20  _{-0.03} ^{+0.04}$                           &$0.205  \pm0.085$	  \\
		$	\Omega_{bc}$&$0.35  \pm0.06 $                      &$0.361  \pm0.112$	\\
		$	\Omega_{bb}$&$0.75  _{-0.09} ^{+0.08} $                     &$0.850  \pm0.160$	 \\
		$	\Xi^{\prime}_{bc}$&$0.36  \pm0.06 $                       & $0.187  \pm0.05$	  \\
		$	\Omega^{\prime}_{bc}$&$0.38  _{-0.07} ^{+0.06} $  & $0.255  \pm0.125$	 \\
		
		\hline\hline
	\end{tabular}%
	\caption{The residues of $2S$ excitation of the doubly heavy baryons in $\mathrm{GeV^3}$ and comparison of the results with other predictions.}
	\label{tab6}
\end{table}

At  the end of this section, it is instructive to give the contributions of perturbative and each nonperturbative operators in the OPE side of the calculations for the $\Xi_{cc}$ state and $\not\!q$ structure as an example. These contributions are depicted in table~\ref{tab:contribution}. As is seen from this table, the main contribution comes from the perturbative part and we see a good convergence of the OPE: the higher the dimension of the nonperturbative operator, the lower its contribution; and all the requirements of the method discussed above are satisfied. 

\begin{table}[tbp]
	\centering
	\begin{tabular}{|c|c|c|c|c|c|c|c|c|c|}
		
		\hline Operator& Perturbative & \, \,\,\,3d\, \,\,\,  &  \, \,\,\, 4d \, \,\,\,& \, \,\,\,5d\, \,\,\, & \, \,\,\, 6d \, \,\,\,& \, \,\,\,7d \, \,\,\,&\, \,\,\, 8d \, \,\,\,& \, \,\,\,9d \, \,\,\, & \, \,\,\, 10d  \, \,\,\, \\ \hline\hline
		Contribution& $69.4 \% $     &$20.5 \%$   &$4.0 \%$&$0 \%$&	$2.8\% $&$0\%$&$1.6\%$ & $1.0\%$ & $0.7\%$	\\  
	
		\hline\hline
	\end{tabular}%
	\caption{Contributions of perturbative and different nonperturbative operators in the OPE side for the $\Xi_{cc}$ state and $\not\!q$ structure in the  average values of auxiliary parameters. Note that 3d$-$10d stand for the nonperturbative operators with three to ten dimensions. }
	\label{tab:contribution}
\end{table}   

\newpage
\section{Summary and conclusion}\label{Sum} 
 The quark model makes predictions about the hadronic states containing single, double and triple heavy quarks. The investigation of the properties of the doubly heavy baryons represents a promising area in particle physics. To study any type of interactions/decays of the doubly heavy baryons, we need the exact values of the masses and residues of these baryons. The values of the spectroscopic parameters calculated from theory can also guide experimental groups to search for the unseen baryon members predicted by the quark model. For this purpose, we determined the spectroscopic parameters of the doubly heavy baryons with a higher accuracy in the ground, first orbitally and first radially excited states. Among the nonperturbative approaches, we used the QCD sum rule formalism as one of the most powerful and predictive methods based on the QCD Lagrangian. In order to increase the accuracy, we included into the analyses the nonperturbative operators up to dimension 10. After fixing the auxiliary parameters, we extracted the masses and residues of the ground, first orbitally and first radially excited states of the doubly heavy baryons as shown in table~\ref{tab:results} and compared our results with the predictions of other approaches and the existing data for the ground $	\Xi_{cc}$ state at different tables. Our result on the mass of the ground $	\Xi_{cc}$ state is in a good consistency with the experimental data of the LHCb collaboration. Our results on the masses are overall consistent with the predictions of other studies except for some channels. Regarding the residue, we have inverse situation: Except for some channels, we see considerable discrepancies of our results with the previous predictions.
 Our predictions may help experimental groups in the course of their search for the unseen members of the doubly heavy baryons. They may also be checked by other nonperturbative approaches in future. As said, in the calculations of any physical quantities related to the interactions/decays of these baryons, the values of the residues are immediately required. Our results that have been provided with higher accuracy can be served as inputs to study the decay properties of the considered doubly heavy baryons in their ground and excited states. Such investigations are required to estimate the width and lifetime of the doubly heavy baryons.
 
\section*{ACKNOWLEDGMENTS}
The authors thank H. R. Moshfegh for his useful discussions. K. Azizi is thankful to Iran National Science Foundation (INSF)
for the partial financial support provided under the elites Grant No. 4025036.

\section*{APPENDIX: EXPRESSIONS OF SPECTRAL DENSITIES}
In this appendix, we present the explicit expressions of the spectral densities obtained from the calculations:

 \begin{align}
 	\rho^{S(pert)}_{1}(s)&=\frac{A}{128 \pi^4}	\int_{0}^{1} \, du \int_{0}^{1-u}
  dv \,  \frac {-3  \,  D_1  \,  \Theta(D_1) } {Z_1^4 \, Z_3^2} \nonumber\\
  &\Bigg\{ \Bigg(-2 \, m_{Q} \, u \, Z1^2 \, (m_{Q'} \, v - 3 \,  m_q \, Z_2) \, Z_3 -  v \, Z_2 \, \Big(-6 \, m_{Q'} \, m_q \, Z_1^2 \, Z_3 + 5 \, u \, (3 \, D_1 \, Z_1^2 + 2 \, s  \, u  \, v \, Z_2 \, Z_3) \Big) \Bigg)  \nonumber\\
 	 &+2 \, t \, u \,  v \,  \Bigg(-3 \, D_1 \, Z_1^2 \, Z_2 - 2 (-m_{Q} \, m_{Q'} \, Z_1^2 + s \, u \, v \, Z_2^2) Z_3 \Bigg)\nonumber\\
 	 &-  t^2 \Bigg(-2 \, m_{Q} \, u \, Z_1^2 \, (-m_{Q'} \, v - 3 \, m_q \, Z_2) \, Z_3 - 
 	 v \, Z_2 \, \Big(-6 \, m_{Q'} \, m_q \, Z_1^2 \,  Z_3 - 5 \, u (3 \, D_1 \, Z_1^2 + 2 \, s \, u \, v \, Z_2 \, Z_3) \Big) \Bigg) \Bigg\},
 \end{align}
 
  \begin{align}
 	\rho^{S(pert)}_{2}(s)&=\frac{A}{64 \pi^4}	\int_{0}^{1} \, du \int_{0}^{1-u}
 	dv \,  \frac {3  \,  D_1  \,  \Theta(D_1) } {Z_1^3 \, Z_3^2} \nonumber\\
 	&\Bigg\{ \Bigg( D_1 \, Z_1^2 \, \Big(3 (m_{Q} \, u + m_{Q'} \, v) - m_q \, Z_2 \Big) - \Big(-s \, u \, v \, Z_2  \, (3 \, m_{Q}  \, u - m_q \, Z_2) + 
 	m_{Q'} \, (5 \, m_{Q} \, m_q \, Z_1^2 - 3 \, s \, u \, v^2 \, Z_2) \Big) \, Z_3 \Bigg) \nonumber\\
 	&-2 \, t \, m_q \Bigg(-D_1 \, Z_1^2 \,  Z_2 - (-m_{Q'} \, m_{Q} \, Z_1^2 + s \, u \, v \, Z_2^2) \, Z_3 \Bigg)\nonumber\\
 	&+ t^2   \, \Bigg(D_1 \, Z_1^2 \, \Big(-3 \, (m_{Q} \, u + m_{Q'} \,  v) - m_q  \, Z_2 \Big) - \Big(-s \, u \, v  \,Z_2 \, (-3 \, m_{Q} \, u - m_q \, Z_2) + m_{Q'} \, (5 \, m_{Q} \, m_q \, Z_1^2 \nonumber\\
 	&+ 3 \, s \, u \, v^2 \, Z_2) \Big) \, Z_3 \Bigg) \Bigg\},
 \end{align}

\begin{align}
	\rho^{A(pert)}_{1}(s)&=\frac{1}{256 \pi^4}	\int_{0}^{1} \, du \int_{0}^{1-u}
	dv \,  \frac { D_1   \,  \Theta(D_1) } {Z_1^4 \, Z_3^2}\nonumber\\
	&\Bigg\{\Bigg(-2 \, m_{Q} \, u \, Z_1^2 \, (-13 \, m_{Q'} \, v - m_q  \, Z_2) \, Z_3 - v \, Z_2 \Big(-2 \, m_{Q'} \, m_q  \, Z_1^2 \, Z_3 + 15 \, u (3 \, D_1 \, Z_1^2 + 2 \, s \, u \, v \, Z_2 \, Z_3)\Big) \Bigg) \nonumber\\
	&-2 \, t \, \Bigg(-2  \, m_{Q}  \, u  \, Z_1^2 \,  (m_{Q'} \,  v - 2 \,  m_q  \, Z_2)  \, Z_3 - 
	v \,  Z_2 (9  \, D_1  \, u  \, Z_1^2 - 4  \, m_{Q'}  \, m_q  \, Z_1^2  \, Z_3 + 6 \,  s  \, u^2  \, v \,  Z_2  \, Z_3) \Bigg)\nonumber\\
	& + t^2 \,  \Bigg(-2  \, m_{Q} \,  u  \, Z_1^2 (-11  \, m_{Q'} \,  v - 5  \, m_q \,  Z_2)  \, Z_3 - 
	5  \, v  \, Z_2 \Big(-2 \,  m_{Q'}  \, m_q  \, Z_1^2 \,  Z_3 - 3  \, u (3 \,  D_1  \, Z_1^2 + 2 \,  s  \, u  \, v  \, Z_2  \, Z_3)\Big)\Bigg)\Bigg\},
\end{align}

\begin{align}
	\rho^{A(pert)}_{2}(s)&=\frac{1}{128 \pi^4}	\int_{0}^{1}  \, du \int_{0}^{1-u}
	dv \,  \frac { D_1  \,  \Theta(D_1) } {Z_1^3 \, Z_3^2}\nonumber\\
	&\Bigg\{ \Bigg(-D_1   \, Z_1^2   \, (-m_{Q}   \, u - m_{Q'}   \, v - 13   \, m_q   \, Z_2) - \Big(s  \,  u   \, v   \, Z_2 (-m_{Q}  \,  u - 13   \, m_q   \, Z_2) + m_{Q'}   \, (15   \, m_{Q}   \, m_q   \, Z_1^2 - s  \,  u  \,  v^2   \, Z_2)\Big) Z_3 \Bigg)\nonumber\\
	&+ 2 \, t  \Bigg(D_1  \, Z_1^2 \Big(2  \, (m_{Q}  \, u + m_{Q'} \,  v) - m_q  \, Z_2 \Big) - \Big(-s  \, u  \, v  \, Z_2 \,  (2 \,  m_{Q}  \, u - m_q  \, Z_2) + 
	m_{Q'} (3  \, m_{Q}  \, m_q  \, Z_1^2 - 2  \, s \,  u  \, v^2  \, Z_2)\Big) \,  Z_3\Bigg)\nonumber\\
	&+t^2 \, \Bigg(D_1   \, Z_1^2 \Big(-5 (m_{Q}  \,  u + m_{Q'}   \, v) - 
	11  \,  m_q   \, Z_2 \Big) - \Big(-s  \,  u   \, v  \,  Z_2 (-5   \, m_{Q}   \, u - 11   \, m_q   \, Z_2) + 
	5   \, m_{Q'}   \, (3   \, m_{Q}   \, m_q   \, Z_1^2 \nonumber\\
	&+ s   \, u   \, v^2   \, Z_2)\Big) Z_3 \Bigg) \Bigg\},
\end{align}

 \begin{align}
	\rho^{S(3d)}_{1}(s)&=\frac{3A}{32 \pi^4}	\int_{0}^{1} \, du \int_{0}^{1-u}
	dv \,    \,  \langle \overline{q} q \rangle \,  \Theta(D_2) \Bigg\{ 2 \, m_{Q} \, u + (2 \, m_{Q'}  + 5 \, m_q \, u) \, Z_4 +2 \, t \, ( m_q  u Z_4) + t^2\, \Big(-2  \, m_{Q}   \, u \nonumber\\
	&- (2  \, m_{Q'}   -5 \, m_q  \, u )Z_4 \Big) \Bigg\},&&&&&&&&&
\end{align}

\begin{align}
	\rho^{S(3d)}_{2}(s)&=\frac{A}{16 \pi^4}	\int_{0}^{1} \, du \int_{0}^{1-u}
	dv \, \langle \overline{q} q \rangle \,  \Theta(D_2)  \Bigg\{  - \Big(2  \, D_2 + 5  \, m_{Q'}   \, m_{Q}  + u  \, (3 m_{Q}   \, m_q + s - s \,  u) + 3  \, m_{Q'}   \, m_q  \, Z_4  \Big) \nonumber\\
	&+2 \, t \, (2   \, D_2 - m_{Q'}   \, m_{Q} + s  \,  u  \, Z_4) + t^2\, \Big(-2   \, D_2 - 5   \, m_{Q'}   \, m_{Q} + 3   \, m_{Q}   \, m_q   \, u + (3 m_{Q'}   \, m_q  - s  \,  u )Z_4 \Big) \Bigg\},&&&&&&&&&
\end{align}

\begin{align}
	\rho^{A(3d)}_{1}(s)&=\frac{1}{192 \pi^4}	\int_{0}^{1} \, du \int_{0}^{1-u}
	dv \,  \langle \overline{q} q \rangle \,  \Theta(D_2)  \Bigg\{   \Big(2  \,  m_{Q}  \,  u + (2   \, m_{Q'} + 45   \, m_q  \,  u )Z_4\Big) \nonumber\\
	&+2 \, t \, \Big(4  \, m_{Q}   \, u + (4 \,  m_{Q'}   + 9  \, m_q  \, u) Z_4 \Big) + 5 \, t^2\, \Big(-2   \, m_{Q}   \,  u - (2   \, m_{Q'}   - 9   \, m_q  \,  u )Z_4 \Big) \Bigg\},&&&&&&&&&&&&&&&&&&&&&&&&&&&&&&&&&
\end{align}

\begin{align}
	\rho^{A(3d)}_{2}(s)&=\frac{1}{96 \pi^4}	\int_{0}^{1} \, du \int_{0}^{1-u}
	dv \,   \langle \overline{q} q \rangle \,  \Theta(D_2) \Bigg\{   \Big(26 \, D_2 - m_{Q'}  (15  \,  m_{Q}  + m_q - m_q  \,  u) - u (m_{Q}   \,  m_q - 13  \,  s  \,  Z_4) \Big) \nonumber\\
	&-2 \, t \, \Big(2   \,  D_2 + 3  \,   m_{Q'}  \,   m_{Q} + u   \,  (2 m_{Q}  \,   m_q + s - s   \,  u) + 2   \,  m_{Q'}   \,  m_q   \,  Z_4 \Big)  \nonumber\\
	&+ t^2\, \Big(-22   \,  D_2+ 5  \,   m_{Q}    \,  m_q   \,  u + 5   \,  m_{Q'}  (-3   \,  m_{Q}  + m_q - m_q   \,  u) - 11   \,  s _u  \,   Z_4 \Big) \Bigg\},&&&&&&&&&&&&&&&&&&&&&
\end{align}

 \begin{align}
	\rho^{S(4d)}_{1}(s)&=\frac{A}{256 \pi^4}	\int_{0}^{1}  \, du \int_{0}^{1-u}
	dv \,  \Big\langle\frac{\alpha_{s}GG}{\pi}\Big\rangle\ \,  \frac {-3 \, u \, v\,  Z_2   \,  \Theta(D_3) } {Z_1^4}\nonumber\\
	&\Bigg\{\Big(3 u^2 + u (-3 + 7 \, v) - 3 \,  v  \, Z_3 \Big)+ 2 \, t  \Big(u^2 + u (-1 + 3 \, v) - v \, Z_3\Big)
	+t^2 \, \Big(3 u^2 + u (-3 + 7 \, v) - 3 \, v \, Z_3 \Big) \Bigg\},&&&&&&&&&&&&&&&&&&&&&&&&&&&&&&&&&&&&&&&&&&&&&&&&&&&&&
\end{align}

\begin{align}
	\rho^{S(4d)}_{2}(s)&=\frac{A}{128 \pi^4}	\int_{0}^{1}  \, du \int_{0}^{1-u}
	dv \,  \Big\langle\frac{\alpha_{s}GG}{\pi}\Big\rangle\ \,  \frac {    \Theta(D_3) } {Z_1^4}\nonumber\\
	&\Bigg\{ -\Bigg[m_{Q}  \,  u  \, Z_1 \Big(u^2 + u (-1 + 5 \,  v) - 3  \, v  \, Z_3\Big) + 
	v \Big[\Big(m_q  \, u + m_{Q'}  \,  (2 + 8 \,  u) \Big) v^3 - 
	3 \, m_{Q'}  \,  v^4 \nonumber\\
	&- \Bigg( (4  \, m_{Q'}  \,  + m_q) u (-1 + 2 \,  u) v  + \Big(2  \, m_q  \, u + 
	m_{Q'}  \,  (-1 + 11  \, u) \Big) v^2 \Bigg)Z_4 + (3 \,  m_{Q'}  \,  + m_q) u^2  \, Z_4^2 \Big] \Bigg]\nonumber\\
	&- 2 \, t  \Big(m_q \, u \, v  \, Z_1 \, Z_2 \Big)
	+t^2 \, \Bigg[m_{Q} u Z_1 \Big(u^2 + u (-1 + 5 v) - 3 v Z_3 \Big) + 
	v \Bigg(m_q u Z_1 Z_2 + 
	m_{Q'} \Big(2 (1 + 4 \,  u) v^3 \nonumber\\
	&- 3  \, v^4 - \Big(
	4  \, u (-1 + 2  \, u) v   + (-1 + 11  \, u) v^2  \Big)\, Z_4 + 3  \, u^2  \, Z_4^2 \Big)\Bigg) \Bigg] \Bigg\},&&&&&&&&&&&&&&&&&&&&&&&&&&&&&&&&&&&&
\end{align}

\begin{align}
	\rho^{A(4d)}_{1}(s)&=\frac{1}{512 \pi^4}	\int_{0}^{1}  \, du \int_{0}^{1-u}
	dv \,  \Big\langle\frac{\alpha_{s}GG}{\pi}\Big\rangle\ \,  \frac { \, u \, v\,  Z_2   \,  \Theta(D_3) } {Z_1^4}\nonumber\\
	&\Bigg\{- \Big(u^2 - u (1 + 11 \, v) - v \, Z_3 \Big)+ 2 \, t  \Big(u^2 + u (-1 + 7 \, v) - v \, Z_3\Big)
	-t^2 \, \Big(u^2 - u (1 + 11 \, v) - v \, Z_3\Big) \Bigg\},&&&&&&&&&&&&&&&&&&&&&&&&&&&&&&&&&&&&&&&&&&&&&&&&&&&&&
\end{align}

\begin{align}
	\rho^{A(4d)}_{2}(s)&=\frac{1}{2304 \pi^4}	\int_{0}^{1}  \, du \int_{0}^{1-u}
	dv \,  \Big\langle\frac{\alpha_{s}GG}{\pi}\Big\rangle\ \,  \frac {    \Theta(D_3) } {Z_1^4}\nonumber\\
	&\Bigg\{ \Bigg[m_{Q}   \, u (19  \,  u^4 + 2  \,  u^3 (-17 + 8   \, v) - u^2 (-15 + v) Z_3 + 12   \, u   \, v^2   \, Z_3 - 
	15  \,  v^2   \, Z_3^2) + 
	v \Bigg(33   \, m_q   \, u   \, Z_1   \, Z_2 \nonumber\\
	&+ m_{Q'} \Big(2 (-17 + 8  \,  u) v^3 + 19   \, v^4 + 12  \,  u^2   \, v   \, Z_4 - (-15 + u) v^2   \, Z_4 - 
	15   \, u^2   \, Z_4^2\Big)\Bigg) \Bigg]\nonumber\\
	& +2 \, t  \Bigg[2  \, m_{Q} \,  u (u^4 + 2 \,  u^3  \, Z_3 - 6  \, u \,  v^2  \, Z_3 - u^2 (3 + v) Z_3 + 3  \, v^2  \, Z_3^2) + 
	v \Bigg(3  \, m_q \,  u  \, Z_1  \, Z_2 + 
	2 m_{Q'} \Big(v^4 - 6  \, u^2  \, v Z_4  \nonumber\\
	&- (3 + u) v^2 \,  Z_4 + 2  \, v^3  \, Z_4 + 3 \,  u^2 \,  Z_4^2\Big)\Bigg) \Bigg]
	+t^2 \, \Bigg[m_{Q}   \, u (-23   \, u^4 + u^3 (26 - 8   \, v) + u^2 (-3 + (8 - 5  \,  v) v) \nonumber\\
	&+ 
	12   \, u   \, v^2   \, Z_3 + 3   \, v^2   \, Z_3^2) + 
	v \Bigg(-39  \,  m_q   \, u   \, Z_1   \, Z_2 + 
	m_{Q'} \Big(2 (13 - 4  \,  u) v^3 - 23  \,  v^4 + 12   \, u^2   \, v   \, Z_4 + (-3 + 5   \, u) v^2   \, Z_4 \nonumber\\
	&+ 
	3   \, u^2   \, Z4^2\Big)\Bigg) \Bigg] \Bigg\},&&&&&&&&&&&&&&&&&&&&&&&&&&&&&&&&&&&&&&&&&&&&&&&&&&&&&
\end{align}

 \begin{align}
	\rho^{S(5d)}_{1}(s)=0,&&&&&&&&&&&&&&&&&&&&&&&&&&&&&&&&&&&&&&&&&&&&&&&&&&&&&&&&&&&&&&&&&&&&&&&&&&&&&&&&&&&&&&&&&&&&&&&&&&&&&&&&&&&&&&&&&&&&&&&&&&
\end{align}

\begin{align}
	\rho^{S(5d)}_{2}(s)&=\frac{5A}{64 \pi^4}	\int_{0}^{1} \, du \int_{0}^{1-u}
	dv \,   \, \langle \overline{q}g_s\sigma Gq\rangle \, u \, Z_4   \,  \Theta(D_4)  \Big(1-2 \, t+ t^2 \Big),&&&&&&&&&&&&&&&&&&&&&&&&&&&&&&&&&&&&&&&&&&&&&&&&&&&&&&&&&&&&
\end{align}

\begin{align}
	 \rho^{A(5d)}_{1}(s)=0,&&&&&&&&&&&&&&&&&&&&&&&&&&&&&&&&&&&&&&&&&&&&&&&&&&&&&&&&&&&&&&&&&&&&&&&&&&&&&&&&&&&&&&&&&&&&&&&&&&&&&&&&&&&&&&&&&&&&&&&&&&
\end{align}

\begin{align}
	\rho^{A(5d)}_{2}(s)&=\frac{5}{384 \pi^4}	\int_{0}^{1} \, du \int_{0}^{1-u}
	dv \,    \, \langle \overline{q}g_s\sigma Gq\rangle \, u \, Z_4   \,  \Theta(D_4) \Big(-13+2 \, t+ 11 \, t^2 \Big),&&&&&&&&&&&&&&&&&&&&&&&&
\end{align}

where

\begin{eqnarray}
		D_1&=&-\frac{Z_3}{Z_1^2} \Big((m_{Q}^2 \, u + m_{Q'}^2 \, v)Z_1 + s \, u \, v \, Z_2 \Big),\nonumber\\
		D_2&=&-m_{Q}^2 u - (m_{Q'}^2  - s u )Z_4,\nonumber\\
		D_3&=&\frac{Z_3}{Z_1^2} \Big((m_{Q}^2 \, u  + m_{Q'}^2 \, v) Z_1 + s \, u \, v \, Z_2 \Big) ,\nonumber\\
		D_4&=&-m_{Q}^2 \,  u - (m_{Q'}^2  - s \,  u )Z_4,
\end{eqnarray}

and we have defined

\begin{eqnarray}
Z_1&=& u^2 + u (-1 + v) + (-1 + v) \, v,\nonumber\\
Z_2&=&1 - u - v,\nonumber\\
Z_3&=&1 - v,\nonumber\\
Z_4&=&1 -u.
\end{eqnarray}


\end{document}